\documentclass[apl,10pt,twocolumn,prb,noshowpacs,superscriptaddress,nofootinbib]{revtex4-2}

\usepackage{graphicx}
\usepackage{multirow}
\usepackage{color}
\usepackage{amsmath}
\usepackage{bm}
\usepackage[dvipsnames]{xcolor}
\usepackage{braket}
\usepackage{siunitx}
\usepackage[hidelinks]{hyperref}
\usepackage{hyperxmp}
\usepackage{booktabs}
\usepackage{array}
\usepackage{subfigure}
\usepackage{mhchem}
\usepackage{hhline}
\usepackage{placeins}

\renewcommand{\vec}[1]{\boldsymbol{#1}}

\newcommand{\appropto}{\mathrel{\vcenter{
  \offinterlineskip\halign{\hfil$##$\cr
    \sim\cr\noalign{\kern0.1pt}\propto\cr\noalign{\kern-0.1pt}}}}}

\newcommand{\ketdeco}[1]{\textcolor{darkgray}{\ket{#1}}}
\newcommand{\bradeco}[1]{\textcolor{darkgray}{\bra{#1}}}

\newcommand{\spindensity}{magnetic electron density}

\usepackage[
    type={CC},
    modifier={by},
    version={3.0},
]{doclicense}

\begin{document}

\title{Probing the real-space density of spin-entangled electrons}

\author{Federico Pisani}
\thanks{These authors contributed equally to this work.}
\affiliation{Laboratory for Quantum Magnetism, Institute of Physics, Ecole Polytechnique Fédérale de Lausanne (EPFL), CH-1015 Lausanne, Switzerland}

\author{Leonie Spitz}
\thanks{These authors contributed equally to this work.}
\affiliation{Quantum Criticality and Dynamics Group, Paul Scherrer Institute, CH-5232 Villigen-PSI, Switzerland}
\affiliation{Institute for Quantum Electronics, ETH Zurich, CH-8093 H\"onggerberg, Switzerland}

\author{Libor Voj\'{a}\v{c}ek}
\thanks{These authors contributed equally to this work.}
\affiliation{PSI Center for Scientific Computing, Theory and Data, Paul Scherrer Institute, CH-5232 Villigen-PSI, Switzerland}
\affiliation{National Centre for Computational Design and Discovery of Novel Materials (MARVEL), CH-5232 Villigen PSI, Switzerland}

\author{Flaviano José dos Santos}
\affiliation{PSI Center for Scientific Computing, Theory and Data, Paul Scherrer Institute, CH-5232 Villigen-PSI, Switzerland}
\affiliation{National Centre for Computational Design and Discovery of Novel Materials (MARVEL), CH-5232 Villigen PSI, Switzerland}
\affiliation{Brazilian Center For Research in Physics (CBPF), Rua Dr. Xavier Sigaud 150, Urca, Rio de Janeiro - RJ, 22290-180, Brazil}

\author{Alberto Carta}
\affiliation{PSI Center for Scientific Computing, Theory and Data, Paul Scherrer Institute, CH-5232 Villigen-PSI, Switzerland}
\affiliation{National Centre for Computational Design and Discovery of Novel Materials (MARVEL), CH-5232 Villigen PSI, Switzerland}

\author{Bastien Dalla Piazza}
\affiliation{Institute of Physics, Ecole Polytechnique Fédérale de Lausanne (EPFL), CH-1015 Lausanne, Switzerland}

\author{Stanislav E. Nikitin}
\affiliation{PSI Center for Neutron and Muon Sciences, Paul Scherrer Institut, CH-5232 Villigen-PSI, Switzerland}

\author{Karl W. Krämer}
\affiliation{PSI Center for Neutron and Muon Sciences, Paul Scherrer Institut, CH-5232 Villigen-PSI, Switzerland}
\affiliation{Department of Chemistry, Biochemistry, and Pharmaceutical Sciences, University of Bern, CH-3012 Bern, Switzerland}

\author{Björn Fåk} 
\thanks{Deceased. We acknowledge Björn’s contributions to this work, in particular his support as instrument scientist during the experiment.}
\affiliation{Institut Laue-Langevin, 71 avenue des Martyrs, CS 20156, 38042 Grenoble cedex 9, France}

\author{Taro Nakajima} 
\affiliation{Institute for Solid State Physics, The University of Tokyo, Kashiwa 277-8581, Japan}
\affiliation{Institute of Materials Structure Science, High Energy Accelerator Research Organization, Tsukuba, Ibaraki 305-0801, Japan}
\affiliation{RIKEN Center for Emergent Matter Science, Wako 351-0198, Japan}

\author{Daichi Ueta} 
\affiliation{Institute of Materials Structure Science, High Energy Accelerator Research Organization, Tsukuba, Ibaraki 305-0801, Japan}

\author{Hiraku Saito}
\affiliation{Institute for Solid State Physics, The University of Tokyo, Kashiwa 277-8581, Japan}

\author{Jian-Rui Soh}
\affiliation{Quantum Innovation Centre (Q.InC), Agency for Science, Technology and Research (A*STAR), 2 Fusionopolis Way, Innovis \#08-03, Singapore 138634, Singapore}
\affiliation{Centre for Quantum Technologies, National University of Singapore, 3 Science Drive 2, Singapore 117543, Singapore}

\author{Nicola Marzari}
\affiliation{PSI Center for Scientific Computing, Theory and Data, Paul Scherrer Institute, CH-5232 Villigen-PSI, Switzerland}
\affiliation{Theory and Simulation of Materials (THEOS), and National Centre for Computational Design and Discovery of Novel Materials (MARVEL), École Polytechnique Fédérale de Lausanne (EPFL), CH-1015 Lausanne, Switzerland}
\affiliation{U Bremen Excellence Chair, Bremen Center for Computational Materials Science, MAPEX Center for Materials and Processes, University of Bremen, D-28359 Bremen, Germany}

\author{Henrik M. Rønnow}
\affiliation{Laboratory for Quantum Magnetism, Institute of Physics, Ecole Polytechnique Fédérale de Lausanne (EPFL), CH-1015 Lausanne, Switzerland}

\begin{abstract}
On the textbook example of an isolated antiferromagnetic Heisenberg dimer, we demonstrate that the magnetic form factor and the \spindensity{} distribution can be extracted from the momentum-dependence of the inelastic neutron scattering (INS) intensity of a magnetic excitation. We measure the three-dimensional (3D) magnetic structure factor of the singlet-to-triplet excitation in Cu(II) acetate monohydrate with INS. Using a minimal parametrization of the \spindensity{}, we deduce the real-space density of the spin-entangled electrons and the transfer of \spindensity{} between metal and ligand atoms from the experimental data. Density functional theory (DFT) calculations reproduce the measured structure factor quantitatively, providing a direct validation of DFT broken-symmetry spin densities against full 3D INS data. The quantitative agreement between experiment, parametrization, and theory establishes a robust framework for determining magnetic form factors and the \spindensity{} in a broad range of magnetic materials and demonstrates INS as a probe of the envelope of spatial electronic wavefunctions.
\end{abstract}

\maketitle

\section{Introduction}

\begin{figure*}[t]
\includegraphics[width=\linewidth]{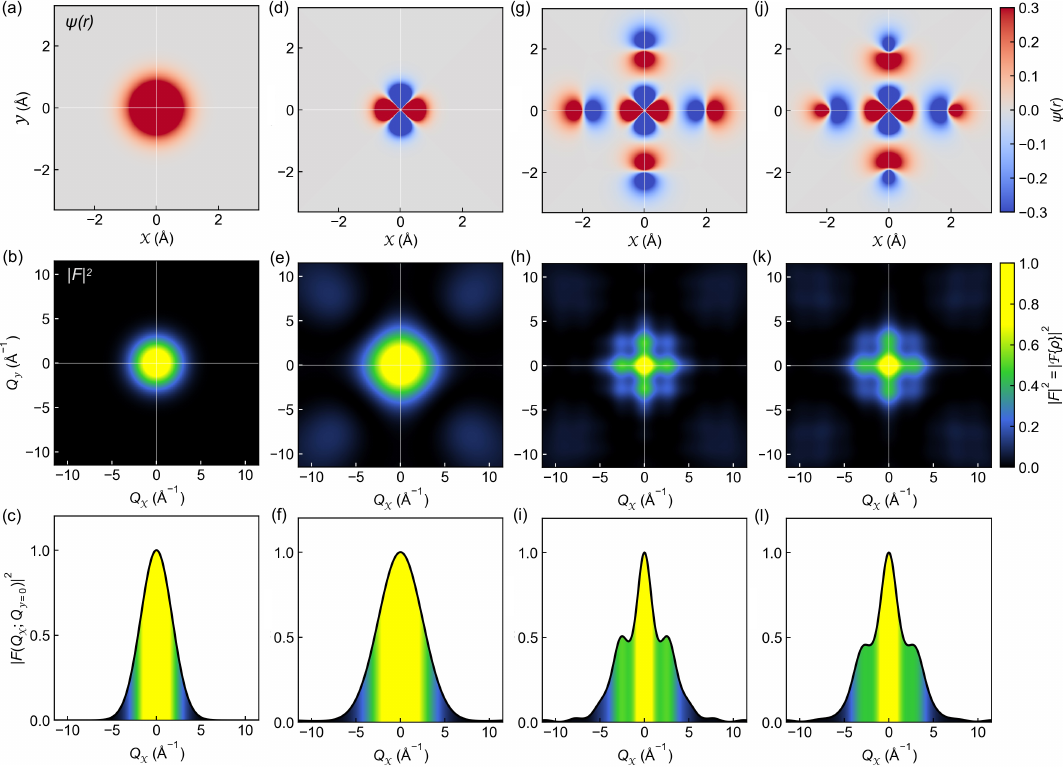}
\caption{Systematic analysis of the impact of the shape of a spatial electronic wavefunction on the shape of the magnetic form factor in momentum space for a $s$ orbital (a-c), the $d_{\mathcal{X}^2-\mathcal{Y}^2}$ orbital (d-f),  $3d$-$2p$ hybrid orbitals in a plaquette geometry (g-i), and $3d$ - $2sp$ hybrid orbitals in a plaquette geometry (j-l). The upper row shows a two-dimensional (2D) projection of the wavefunction  in real space. The middle row shows the  corresponding absolute square of the magnetic structure factor in reciprocal space, which is given by the absolute square of the Fourier transform of the \spindensity{} $\rho = |\psi|^2$. The bottom row shows a one-dimensional (1D) cut along $Q_{\mathcal{X}}$ at $Q_{\mathcal{Y}} = 0$. The metal-ligand distance in panels (j-l) is $\qty{1.97}{\AA}$, which corresponds to the average Cu-O distance in a \ce{CuO4} unit of Cu(II) acetate monohydrate.}
\label{fig:intro}
\end{figure*}

Quantum magnetic insulators host a remarkably diverse range of magnetic phases and phenomena spanning from relatively conventional behavior to highly exotic states \cite{sachdev_quantum_2008}. Some develop long-ranged magnetic order, others remain disordered even at lowest temperatures, and some display a complex coexistence of both order and disorder \cite{lee_end_2008,vasiliev_milestones_2018,manousakis_spin-_1991}. The primary tools for capturing the magnetic ground state and the excitation spectra of quantum phases are spin Hamiltonians, which encode the mutual interactions among spins in the underlying many body systems. They often provide an excellent description of the macroscopic many-body states realized in real materials, but in some cases, e.g. in the presence of non-negligible quantum fluctuations, they can fail to reproduce the experimental ordering behavior of the magnetic moments precisely \cite{Anderson_1951,manousakis_spin-_1991}. Achieving a deeper understanding of why one particular compound develops long-ranged magnetic order while in another fluctuations dominate, calls for moving beyond this effective description. This entails reducing the level of abstraction and examining the microscopic observables behind the model. While the spin operators $\hat{S}$ in the model Hamiltonians stand for effective spins localized on specific lattice sites, the magnetic interactions between them originate from the delocalization of the electron spins between the atoms. Their foundation lies in the distribution of charge density over the magnetic orbitals which we refer to as the \spindensity{}. In polarized magnetic phases, this corresponds to the spin density. The \spindensity{} distribution $\rho(\boldsymbol{r}) = |\psi(\boldsymbol{r})|^2$ is determined by the spatial parts of the electronic wavefunctions $\psi(\boldsymbol{r})$ linking the fundamental electronic properties and the macroscopic magnetic state of correlated electron systems. The connection between microscopic electronic properties, the \spindensity{} distribution and the macroscopic spin Hamiltonian calls for methods to characterize the \spindensity{} with qualitative and quantitative accuracy and promises answers to the question why one material hosts an exotic quantum many body state, while another is a conventional magnet beyond the classic geometry arguments. Apart from the established problems of quantum magnetism, determining the \spindensity{} distribution recently gained relevance in the study of altermagnets which carry a zero net magnetization -- like antiferromagnets -- and are characterized by the spatial separation of orbitals with opposite spin resulting in the presence of an anisotropy of the local spin density \cite{jungwirth_altermagnets_2026}. 
 
The neutron magnetic form factor, $F(\boldsymbol{Q})$, is related to the real-space distribution of the \spindensity{} by a Fourier transform. Hence, the momentum ($\boldsymbol{Q}$)-space distribution of neutron scattering intensities $I(\boldsymbol{Q}) \propto |F(\boldsymbol{Q})|^2$ contains quantitative information about the 3D real-space distribution of the charge density in magnetic orbitals and consequently about the envelope of the underlying electronic wavefunctions. An established method to measure the magnetic form factor with neutrons involves polarized neutron diffraction on nuclear Bragg peaks, where interference between nuclear and magnetic scattering enables extraction of the magnetic scattering amplitude. While this is a property intrinsic to ferromagnets and ferrimagnets, application of a magnetic field to induce a net magnetic moment extends the utility of the method to many systems lacking spontaneous magnetization. Because the magnetic form factor is determined completely, including its phase, the spin-density distribution can then be extracted by a Fourier transform \cite{boothroyd2020}. 

In our work, we show that the dependence of the magnetic form factor on the scattering vector $\boldsymbol{Q}$ can also be measured directly by INS on the magnetic excitations of a spin system and that the scattered intensities contain information about the \spindensity{} distribution of the excited state. This method does not require the presence of a net magnetic moment and is therefore convenient for antiferromagnets. The loss of the phase of the magnetic form factor in this approach is compensated for by theoretical methods. Because the $\boldsymbol{Q}$-dependence of INS intensities of magnetic excitations is not only determined by the $\boldsymbol{Q}$-dependence of the magnetic form factor but also by the dispersion of the excitation, we provide the proof of principle on a very simple system, the zero-dimensional isolated dimer of two spins on two atomic sites, before discussing the generalization of the concept to more complex materials.

The antiferromagnetic Heisenberg dimer is described by the spin Hamiltonian
\begin{equation}
    \label{eq:Heisenberg_Hamiltonian_definition}
    \hat{H} = -2J\hat{\boldsymbol{S}}_1\cdot\hat{\boldsymbol{S}}_2,
\end{equation}
where $J$ is the magnetic exchange coupling constant. For antiferromagnetic coupling, the ground state is the nonmagnetic singlet, the simplest example of an entangled state. The only excitation of the system is a sharp, dispersionless, paramagnetic triplet which is separated from the ground state by a gap of size $\Delta = |2J|$. 

Cu(II) acetate monohydrate is a prototypical isolated Heisenberg dimer material with a spin excitation gap $\Delta =\qty{37}{meV}$ that neither shows significant temperature-dependence nor is affected by deuteration~\cite{guedel_direct_1979,gudel_inelastic_1980}. The magnetic susceptibility follows the Bleaney-Bowers equation for isolated dimers~\cite{bleaney_anomalous_1952,elmali_magnetic_2000}. A previous DFT study confirmed that the magnetic exchange is dominated by a superexchange process mediated by the O and C atoms of the acetate groups linking the Cu sites and reproduced the experimental value of $J$~\cite{kyuzou_electronic_2010}.

The \spindensity{} distribution of transition-metal oxides is dominated by hybrid metal-$3d$ and ligand-$2sp$ orbitals. In the plaquette-like arrangements of Cu and ligand atoms common in copper oxides, the $3d^9$ electronic configuration yields a half-filled $d_{\mathcal{X}^2-\mathcal{Y}^2}$ orbital directed towards the surrounding ligands. The hybridization enables superexchange interactions between neighboring metal ions via virtual electron hopping processes. The covalent character of the bonding leads to a substantial delocalization of the \spindensity{} over the entire metal–ligand network. This delocalization has profound consequences for the $\boldsymbol{Q}$-dependent spin–spin correlations probed by inelastic neutron scattering, motivating experimental and theoretical investigations \cite{freltoft_magnetic_1988,walters_effect_2009}.

The effect of the shape of the hybrid wavefunctions on the neutron magnetic form factor $F(\boldsymbol{Q})$ is demonstrated in Fig.~\ref{fig:intro} by comparing the form factor of isolated $s$ and $d$ orbitals with the form factor of a \ce{CuO4} plaquette. The ionic picture, in which the magnetic orbital is isotropic and centred on the \ce{Cu} site, results in an isotropic Gaussian-like magnetic form factor with a width that corresponds to the extent of the assumed orbital (Figs.~\ref{fig:intro}(a-c)). If the unpaired spin is located primarily in a $d$ orbital, the magnetic form factor reflects the anisotropic shape of the orbital (Figs.~\ref{fig:intro}(d-f)). The fine structure of the $d$-orbital which introduces short-length modulations in real space leads to additional contributions to the magnetic form factor at large $Q$-vectors in reciprocal space as visible in Fig.~\ref{fig:intro}(e). Around the origin of reciprocal space it is still dominated by a central peak due to the fact that the magnetic form factor is the Fourier transform of the absolute square of the wavefunction -- a purely positive function with a finite mean. If covalent metal-ligand hybridization is taken into account (Figs.~\ref{fig:intro}(g-i)), the \spindensity{} is delocalized over the hybrid orbital, resulting in an anisotropic magnetic form factor with a multi-peak structure. The relative height of these peaks can be further affected when the hybrid character of the oxygen $2sp$ orbitals is taken into account (Figs.~\ref{fig:intro}(j-l)).
Although the anisotropic $Q$-dependence of $F(\mathbf{Q})$ in insulators with covalent metal-ligand bonding has been pointed out in multiple studies \cite{walters_effect_2009,mazurenko_covalency_2015}, the majority of inelastic neutron scattering experiments still assume the isotropic form factor of the metal ion, here \ce{Cu}$^{2+}$. In our work, we demonstrate methods with which the magnetic form factor can be accounted for more precisely.

A powerful tool for investigating the hybridization of wavefunctions and for predicting the resulting \spindensity{} distribution is DFT. A study of the effect of covalency on the neutron magnetic form factor in cuprates was presented by Mazurenko \textit{et al.}~\cite{mazurenko_covalency_2015}. Using a Wannier function based approach \cite{anderson_new_1959}, the authors show on the quasi-2D quantum spin system \ce{BaCuSi2O6} with one selected 1D cut in reciprocal space at the energy of the magnetic excitation that the magnetic orbital structure determines the $Q$-dependence of the magnetic form factor. However, the study covers a very limited part of the reciprocal space and does not draw quantitative conclusions on the underlying \spindensity{} distribution.

Using the textbook example of an isolated Heisenberg dimer, we demonstrate that the neutron magnetic form factor of a singlet-to-triplet transition can be measured directly by INS. From the experimental data, we reconstruct the \spindensity{} distribution and the underlying spatial electronic wavefunction quantitatively. While previous studies were limited to one dimension only and did not make quantitative statements, our work provides a fully quantitative comparison between \spindensity{} distributions calculated in DFT and the measured INS intensities in 3D in a correlated quantum material. To facilitate a simple reconstruction of the \spindensity{} distribution without resource-intensive calculations, we further introduce an analytical parametrization based on anisotropic Gaussians which describes the neutron magnetic form factor of copper oxides with a high accuracy. This establishes a robust, accessible method for the characterization of the \spindensity{} distribution in quantum magnets and sets a gold standard for extracting neutron magnetic form factors.

The structure of this article is as follows: In Sec.~\ref{sec:MatMeth}, we provide an introduction to the material Cu(II) acetate monohydrate, to our state-of-the-art INS experiments, and to the analytical and numerical methods we use to calculate and interpret the neutron magnetic form factor. In Sec.~\ref{sec:magformfactor}, we compare the results  and retrieve a quantitative description of the \spindensity{} distribution. In Sec.~\ref{sec:excitation}, we propose and parametrize an extended Hubbard dimer model that captures the interactions between the magnetic orbitals of Cu(II) acetate monohydrate faithfully. In Sec.~\ref{sec:discussion}, we discuss the generality and the consequences of our results before drawing a brief conclusion in Sec.~\ref{sec:conclusion}.

\section{Material and Methods}
\label{sec:MatMeth}

\begin{figure}[t]
\includegraphics[width=\linewidth]{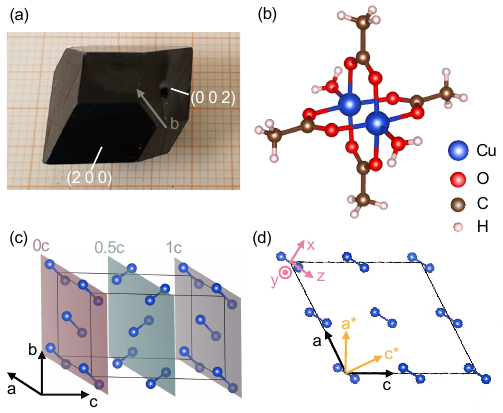}
\caption{(a) Single crystal of deuterated Cu(II) acetate monohydrate used in INS experiments. The facettes perpendicular to the $H$ and $L$ directions are normal to the $(2 0 0)$ and $(0 0 2)$ faces. The $\hat{b}$-axis is parallel to the edge between these facettes, as indicated in gray. (b) Atomic structure of the Cu(II) acetate monohydrate molecule. (c) Projection showing the two orientations of the Cu-Cu dimers in the monoclinic unit cell. The blue connector between Cu ions represents the orientation of the vector between them, not a chemical bond. Planes  indicate the position of the dimers along the $\hat{c}$-axis. (d) Projection of the Cu-Cu dimers on the $ac$ plane of the monoclinic unit cell. Orange arrows indicate the directions of the in-plane reciprocal-lattice vectors. The out-of-plane vector, $\hat{b}^{*}$, is parallel to $\hat{b}$. Pink arrows indicate the Cartesian coordinate system used in the discussion of the data.}
\label{fig:dimer}
\end{figure}

\subsection{Material}

The single crystal of deuterated Cu(II) acetate hydrate \ce{Cu2(CD3COO)4(D2O)2} used in INS experiments is shown in Fig.~\ref{fig:dimer}(a). The crystal growth is described in the Supplemental Material~\cite{supplemental}. \ce{Cu2(CD3COO)4(D2O)2} crystallizes in the inversion-symmetric monoclinic space group C2/c (no.~15). We used single-crystal X-ray diffraction to determine the lattice parameters at room temperature to be $\text{a} = \qty{13.1722(2)}{\AA}$, $\text{b} = \qty{8.5615(1)}{\AA}$, $\text{c} = \qty{13.8610(2)}{\AA}$ and $\beta = \ang{117.026(2)}$ in close agreement with the literature \cite{de_meester_refined_1973}. 
The unit cell hosts four Cu dimers, which are characterized by the short intradimer distance of $\qty{2.61}{\AA}$. The dimer unit is shown in Fig.~\ref{fig:dimer}(b). Each Cu ion is coordinated by four O atoms, forming a \ce{CuO4} plaquette. O and C atoms of the acetate ions bridge two Cu centers. Notably, the Cu centers are related by inversion symmetry, but not by mirror symmetry. As illustrated in Fig.~\ref{fig:dimer}(c), the dimers adopt one of two orientations within the plane perpendicular to the $[101]$ direction, their projections onto the \textit{ac} plane being identical as shown in Fig.~\ref{fig:dimer}(d). Finally, a water molecule is connected to each Cu atom along the Cu–Cu axis, but is positioned outside the dimer. Hydrogen bonds link neighboring molecules, stabilizing the three-dimensional crystal structure. 

When describing the reciprocal space, the plane of interest is defined by the lattice vectors $\mathbf{a}$ and $\mathbf{c}$ (or equivalently $\mathbf{a}^*$ and $\mathbf{c}^*$). For convenience in analyzing the experimental data, we introduce a new coordinate system: $Q_z$ is aligned with the projection of the Cu–Cu dimers onto the $ac$ plane, while $Q_x$ is defined as the direction perpendicular to $Q_z$ within the same plane. The direction out-of the plane, $Q_y$, is taken along $\mathbf{b}$ (or $\mathbf{b}^*$). All relevant vectors are shown in Fig.~\ref{fig:dimer}(c,d). For reference, the components of $Q_x$ and $Q_z$ in reciprocal lattice units (r.l.u.) are $Q_x = [0.9091, 0, 1]$ and $Q_z = [0.9786, 0, -1]$.

\subsection{Inelastic neutron scattering}

For neutron spectroscopy experiments, the deuterated single crystal with mass $\qty{3.45}{\gram}$ shown in Fig.~\ref{fig:dimer}(a) was mounted on an aluminum sample holder equipped with a two-axis goniometer, allowing precise orientation of the crystal. The sample was aligned with the \textit{b}-axis vertical, such that the \textit{ac} plane defined the horizontal scattering plane. The crystal orientation was verified using an X-ray Laue camera and neutron diffraction, the latter also confirming the single-crystallinity of the bulk. 

Preliminary measurements were performed at the High Resolution Chopper Spectrometer (HRC) at the Materials and Life Science Experimental Facility (MLF), J-PARC~\cite{itoh_high_2011,itoh_progress_2019,ueta_sample_2024}. The crystal was sealed in an aluminum can and cooled to a base temperature of $\qty{10}{K}$ in a $\qty{1}{\kelvin}$ refrigerator. Two chopper configurations were employed (i) incident energy $E_{\mathrm{i}} = \qty{100}{meV}$ with chopper frequency $f = \qty{300}{Hz}$, and (ii) $E_\mathrm{i} = \qty{150}{meV}$ with $f=\qty{400}{Hz}$. During this measurement, the sample was rotated through $\ang{180}$ in $\ang{1}$ steps within the neutron beam. The experiment revealed the singlet-to-triplet excitation at approximately $\qty{37}{meV}$ and provided information on its $\boldsymbol{Q}$-dependence, which allowed us to optimize the $Q$-coverage in subsequent experiments. The relevant plots from HRC measurements are shown in the Supplemental Material \cite{supplemental}.

The subsequent INS experiments were performed on the same crystal using PANTHER, the thermal neutron time-of-flight (TOF) spectrometer at the Institut Laue--Langevin (ILL)~\cite{fak_panther_2022}. The sample was placed in an Orange cryostat and data were collected at a base temperature of $\qty{1.5}{\kelvin}$. Empty-can scans were recorded under identical conditions to serve as the background reference. The sample was rotated through $\ang{350}$ in $\ang{1}$ steps during the measurement. An incident energy of $E_\mathrm{i} = \qty{76}{meV}$ and a chopper frequency of $f = \qty{270}{Hz}$ were chosen, with a graphite (004) monochromator. With this configuration, we obtained an energy resolution of \qty{4.5}{meV} at the elastic line and \qty{3.9}{meV} at the singlet-triplet excitation energy. The Horace software was used for visualizing and analyzing the four-dimensional $S(\mathbf{Q}, \omega)$ data~\cite{Ewings_horace_2016}. The single-crystal TOF data were symmetrized with respect to the symmetry operation $C_2$ of the space group.

The datasets display a dispersionless feature centered around $E = \qty[separate-uncertainty]{37.1 \pm 0.3}{meV}$, which can be attributed to the transition from the singlet ground state to the triplet excited state, in agreement with previous studies \cite{guedel_direct_1979}. Because phonon scattering becomes relevant for larger wave vectors $\mathbf{Q}$, we modeled and subtracted the phonon background. Details of the background treatment are presented in the Supplemental Material~\cite{supplemental}.

\subsection{Analytical parametrization}
\label{Analytical model}

We introduce a simple analytical parametrization of the \spindensity{} to describe the neutron cross-section of Cu(II) acetate monohydrate. Cu ions, with a 3$d^9$ electronic configuration, are the primary magnetic centers. In the ionic picture they carry spin~$\frac{1}{2}$. Due to covalent bonding, the Cu \spindensity{} is delocalized onto the ligand O atoms. On the basis of the fit of the parametrization to the experimental data, we gain quantitative insight into the fraction of spin charge migrating on the ligands and on the extent of the spatial wavefunction.

To maintain generality, we begin by introducing the neutron scattering formalism for arbitrary \spindensity{} distributions. The magnetic scattering cross-section for the transition from the ground state $\psi_{\mathrm{i}}$ to a final excited state $\psi_{\mathrm{f}}$ can be expressed as
\begin{equation}
S^{\alpha\alpha}(\vec{Q}) = 
\sum_{\psi_{\mathrm{i}}, \psi_{\mathrm{f}}} 
\left| \left\langle \psi_{\mathrm{f}} \middle| \hat{V}^{\alpha}(\mathbf{Q}) \middle| \psi_{\mathrm{i}} \right\rangle \right|^2,
\quad \alpha = x,y,z,
\end{equation}
where the spin-only neutron scattering operator is defined as
\begin{equation}
    \hat{V}^{\alpha}(\mathbf{Q}) =\sum_j \hat{S}_{j}^{\alpha}  \int \rho_j(\mathbf{r} - \mathbf{R}_j)\, e^{i \mathbf{Q}\cdot \mathbf{r}} \, d\mathbf{r}.
\label{eq:general_V_operator}
\end{equation}
Here, $\rho_j(\mathbf{r})$ is the real-space \spindensity{} distribution associated with spin $S_j$ and centered at $\mathbf{R}_j$, without any assumption about its functional form.
For Cu(II) acetate monohydrate, the relevant spin states are the singlet ground state 
\[
\lvert S \rangle = \frac{1}{\sqrt{2}}
\left(
\lvert \uparrow \downarrow \rangle
-
\lvert \downarrow \uparrow \rangle
\right),
\] 
and the three excited triplet states
\[
\lvert T_+ \rangle = \lvert \uparrow\uparrow \rangle,
\quad
\lvert T_0 \rangle = \frac{1}{\sqrt{2}}\left( \lvert \uparrow\downarrow \rangle + \lvert \downarrow\uparrow \rangle \right),
\quad
\lvert T_- \rangle = \lvert \downarrow\downarrow \rangle.
\]
Considering the two \ce{CuO4} plaquettes forming the dimer, Eq.~\eqref{eq:general_V_operator} becomes
\begin{align} 
\hat{V}^{\alpha}(\mathbf{Q}) &=
\left(
\rho_{\mathrm{Cu_1}}(\mathbf{Q})\, e^{i\mathbf{Q}\cdot\mathbf{R}_{\mathrm{Cu_1}}}
+ \sum_{k=1}^{4} \rho_{\mathrm{O_{k1}}}(\mathbf{Q})\, e^{i\mathbf{Q}\cdot\mathbf{R}_{\mathrm{O_{k1}}}}
\right) \hat{S}_{\mathrm{Cu_1}}^{\alpha} \notag \\[6pt]
&+
\left(
\rho_{\mathrm{Cu_2}}(\mathbf{Q})\, e^{i\mathbf{Q}\,\cdot\mathbf{R}_{\mathrm{Cu_2}}}
+ \sum_{k=1}^{4} \rho_{\mathrm{O_{k2}}}(\mathbf{Q})\,  e^{i\mathbf{Q}\cdot\mathbf{R}_{\mathrm{O_{k2}}}}
\right) \hat{S}_{\mathrm{Cu_2}}^{\alpha}, 
\label{eq:general_total_operator}
\end{align}
where each ${\rho}(\mathbf{Q})$ denotes the Fourier transform of the \spindensity{} $\rho(\mathbf{r})$ associated with the specified atom. Note that only single-spin operators related to the two Cu ions, $\hat{S}_{\mathrm{Cu_1}}^{\alpha}$ and $\hat{S}_{\mathrm{Cu_2}}^{\alpha}$, are used here because Cu atoms are the sole sources of magnetically active electrons.
Because the singlet and triplet states have opposite parity under exchange of the two Cu spins, the matrix elements lead to a relative minus sign between the two \ce{CuO4} units. Rewriting \eqref{eq:general_total_operator} as
\begin{equation}
\hat{V}^{\alpha}(\mathbf{Q}) = V_1(\mathbf{Q})\, \hat{S}_{\mathrm{Cu_1}}^{\alpha}+V_2(\mathbf{Q})\, \hat{S}_{\mathrm{Cu_2}}^{\alpha},
\end{equation}
the magnetic scattering cross-section of one dimer becomes
\begin{equation}
S (\vec{Q}) = \sum_{\alpha = x,y,z} S^{\alpha\alpha}(\vec{Q}) = \frac{3}{4} \left|V_1 (\vec{Q}) - V_2 (\vec{Q})\right|^2.
\end{equation}
In our parametrization, we neglect the small fraction of \spindensity{} that may reside on the C atoms. These atoms are shared nearly symmetrically between the two \ce{CuO4} plaquettes. Consequently, due to the relative minus sign between the two contributions, their net contribution largely cancels out. Any residual asymmetry in the molecular structure may in principle give rise to a small signal in the neutron cross-section. However, the Cu–C distances in the molecule are comparable to the Cu–O separations, making it difficult to disentangle such a weak contribution from the dominant Cu–O signal.

Finally, we need to consider that the system contains two symmetry-inequivalent dimer orientations. Because the two dimers are magnetically isolated, their contributions to the total neutron structure factor can be added. Taking into account a second dimer, $\ce{Cu3}-\ce{Cu4}$, we obtain
\begin{equation}
S (\vec{Q}) = \frac{3}{4} \left( \left|V_1 (\vec{Q}) - V_2 (\vec{Q})\right|^2 + \left|V_3 (\vec{Q}) - V_4 (\vec{Q})\right|^2 \right).
\end{equation}

\begin{table}[t]
\caption{Parameters of the triaxial Gaussians on each atomic site in the analytical parametrization of the \spindensity{}, together with their interpretation and the best-fit values for a voxel of size $dQ = \qty{0.15}{{\AA}^{-1}}$.}
\centering
\renewcommand{\arraystretch}{1.2}
\begin{ruledtabular}
\begin{tabular}{c >{\centering\arraybackslash}p{4.3cm} c}
  
    \textbf{Parameter} & \textbf{Interpretation} & \textbf{Best-fit value} \\
    \midrule
     $A$ & Scaling factor &  \\
    $\displaystyle \frac{\overline{\rho}_{\mathrm{O}}}{\overline{\rho}_{\mathrm{Cu}}+4\,\overline{\rho}_{\mathrm{O}}}$ & Fraction of magnetic electron density on each O site & \qty{5.28(5)}{\%}   \\
    \large $ \sigma_{\mathrm{Cu}\perp}$ & Gaussian width along the Cu--Cu axis & \qty{0.184(1)}{\AA} \\
    \large $\displaystyle \sigma_{\mathrm{Cu}\parallel}$ & Gaussian width within the \ce{CuO4} plaquette & \qty{0.230(1)}{\AA}   \\
    \large $\sigma_{\mathrm{O}\parallel}$ & Gaussian width along the Cu--O bond & \qty{0.427(6)}{\AA} \\
   \large $\sigma_{\mathrm{O}\perp}$ & Gaussian width perpendicular to the Cu--O bond & \qty{0.270(6)}{\AA}  \\

\end{tabular}
\end{ruledtabular}
\label{table:table1}
\end{table}

To obtain closed-form analytic expressions for the functions ${\rho}_j(\mathbf{Q})$, we have to introduce an analytical approximation for the \spindensity{}. From Fig.~\ref{fig:intro}, we infer that the magnetic form factor is dominated by the interference of the contributions from the atomic sites, whereas the contributions of the details of the underlying atomic orbitals are weak. Neutron scattering experiments probe the \spindensity{} only over a finite range of momentum transfer \textbf{Q}, which limits sensitivity to fine real-space features of the underlying atomic orbitals. Deviations from the true atomic-orbital form at very short length scales therefore do not affect the quantities accessible in the experiment. Hence, a good approximation must capture the extent and anisotropy of the orbitals, whereas the fine structure of the orbital shape can be neglected given the limits of the experiment. We parametrize the \spindensity{} on each atomic site as a triaxial Gaussian. Within the restricted \textbf{Q}-range, a triaxial Gaussian provides an accurate and well-behaved representation of the spatial extent and anisotropy of the local \spindensity{}, while allowing an analytical evaluation of its Fourier transform. 

A general Gaussian centered at the origin with principal axes aligned to $(x,y,z)$ is
\begin{equation}
\rho(\mathbf{r}) = \frac{\overline{\rho}}{(2\pi)^{3/2} 
\sigma_x \sigma_y \sigma_z}\,
\exp\!\left(-\frac{x^2}{2\sigma_x^2}
           -\frac{y^2}{2\sigma_y^2}
           -\frac{z^2}{2\sigma_z^2}\right),
\label{eq:real_gaussian}
\end{equation}
where $\overline{\rho}$ is the integrated \spindensity{}, and $\sigma_x$, $\sigma_y$, and $\sigma_z$ are the respective Gaussian widths. The Fourier transform retains the Gaussian form,
\begin{equation}
\rho(\mathbf{Q}) = \overline{\rho} \,
\exp\!\left[-\frac{1}{2}(Q_x^2\sigma_x^2 + Q_y^2\sigma_y^2 + Q_z^2\sigma_z^2)\right].
\label{eq:reciprocal_gaussian}
\end{equation}
The Gaussian parameters are adjusted according to the local geometry of the \ce{CuO4} plaquettes.
For Cu sites, one principal axis is aligned with the Cu–Cu separation; the other two axes lie in the approximately planar \ce{CuO4} environment. For the O sites, the first principal axis points towards the bonded Cu atom, while the remaining two axes lie within the plane perpendicular to the Cu--O axis.
All Gaussian spin densities are inserted into Eq.~\eqref{eq:general_total_operator} using the reciprocal-space expression \eqref{eq:reciprocal_gaussian}. The resulting function, multiplied by a scaling factor $A$ to match the experimental intensity, was fitted with standard least-squares algorithms to the experimental data over a large volume of reciprocal space, with $Q_x \in \left[-5, 5 \right]~\unit{\AA^{-1}}$, $Q_y \in \left[-1, 2 \right]~\unit{\AA^{-1}}$, and  $Q_z \in \left[-5, 5 \right]~\unit{\AA^{-1}}$. Each experimental data point corresponds to the intensity averaged over a small cubic voxel of size $dQ$ in the reciprocal space. The fit yields consistent results for different choices of $dQ$, as reported in Table S2 (Supplemental Material~\cite{supplemental}).
Table~\ref{table:table1} provides a summary of all parameters and the best-fit values for $dQ = \qty{0.15}{{\AA}^{-1}}$. 

\subsection{\textit{Ab initio} calculations}
\label{subsec:DFT_methods}

The electronic structure calculations consist of two parts: (1) a simple yet powerful symmetry-broken collinear DFT+$U$ in Sec.~\ref{sec:magformfactor} to obtain the magnetic form factor and (2) a more exact treatment of the many-body physics that provides the full spectrum of the magnetic orbitals in Sec.~\ref{sec:excitation}.

The DFT calculations in Section~\ref{sec:magformfactor} were performed using \texttt{Quantum ESPRESSO v.7.3}~\cite{giannozzi_quantum_2009, giannozzi_advanced_2017} with the PBEsol functional~\cite{perdew_restoring_2008} and the optimized norm-conserving Vanderbilt pseudopotentials~\cite{Hamann2013_ONCV}.
A Hubbard correction $U=$ 7.3 eV was applied to the Cu 3\textit{d} electrons, calculated using the linear-response theory~\cite{cococcioni_linear_2005} within the density-functional perturbation
theory~\cite{timrov_hubbard_2018} implemented in the \texttt{HP} code~\cite{timrov_hp_2022} (see the Supplemental Material~\cite{supplemental}). A uniform Gamma-centered $2 \times 2 \times 2$ k-point grid was used to sample the Brillouin zone.

The \spindensity{} was directly obtained in the form of a broken-symmetry spin density, where the two Cu atoms within the same molecule were initialized with opposite magnetic moments. The total absolute magnetic moment converged to 2.04~$\mu_{\mathrm{B}}$ per dimer, close to the experimental value~\cite{brown_dinuclear_1973}. See Appendix~\ref{app:singlet_and_DFT} for a detailed discussion of the use of broken-symmetry DFT for open shell systems.

The kinetic energy cut-off for the charge density, $E_{\mathrm{cut},\rho} = 1440$~Ry, ensures a real-space resolution of $\lambda = 2 \pi / k_{\max} = 2 \pi / \sqrt{2 m_{e} E_{\mathrm{cut},\rho} / \hbar^{2}} = 0.043$~$\mathrm{\AA}$ in the spin density Gaussian cube file representation.

The spin density was kept only within muffin-tin spheres around selected sites with a cutoff radius of $1.1$~$\mathrm{\AA}$ for Cu and $0.9$~$\mathrm{\AA}$ for O.

The form factor was then calculated as the Fourier transform of this filtered broken-symmetry spin density. Higher resolution in the reciprocal space was ensured by increasing the real-space data array with zero-padded values to 4 times the dimensions of the original unit cell. All of the post-processing tools were implemented in the publicly available \texttt{fft\_electronic\_spin\_density} Python package~\cite{fft_electronic_spin_density}.

The Hubbard dimer model~\cite{mazurenko_wannier_2007, miyake_screened_2008} we present in Sec.~\ref{sec:excitation} was parametrized with maximally localized Wannier functions (WFs)~\cite{marzari_maximally_1997} representing the 
frontier orbitals (HOMO/LUMO) and computed with non-spin-polarized DFT using norm-conserving pseudopotentials from the ONCV library~\cite{Hamann2013_ONCV} and \texttt{Wannier90}~\cite{pizzi_wannier90_2020} in an energy window between $-0.265$~eV and 0.255~eV with respect to the Fermi level by initially projecting onto one 3$d$ orbital per Cu atom, resulting in 8 WFs for the 4 f.u. unit cell, which hosts 4 dimer molecules. We remark that the resulting frontier WFs extend beyond the initial projections on the Cu atomic centers and contain both the Cu 3\textit{d} and the O 2\textit{sp} character, as shown in Fig.~\ref{fig:Fig5}(c).

Constrained random-phase approximation (cRPA) calculations were performed starting from these frontier WFs using \texttt{RESPACK}~\cite{nakamura_respack_2021} version \texttt{v0201117} with a $2\times2\times2$ \textbf{k}-grid, 288 empty bands, and a convergence sweep over the energy cutoff as shown in the Supplemental Material~\cite{supplemental}.

We compared the exchange coupling parameters from the Hubbard dimer model with the results of the magnetic force theorem (also known as the Liechtenstein-Katsnelson-Antropov-Gubanov, or LKAG, method) implemented in the \texttt{TB2J} package~\cite{he_tb2j_2021}. The LKAG method was applied on WFs derived from DFT+($U = 7.3$~eV). We note that these WFs contain the correlation effects intrinsically  through the effective Hubbard $U$ and they are distinct from the non-magnetic Wannier functions used for the Hubbard dimer model, where the correlation effects enter \textit{a posteriori} through the four-point screened Coulomb parameters $W_{ii}, W_{ij}$, and $J_\mathrm{dir}$.

\section{Magnetic form factor}
\label{sec:magformfactor}

\begin{figure*}[t]
\includegraphics[trim={0.05cm 0 0.1cm 0}, clip, width=1.0\linewidth,]{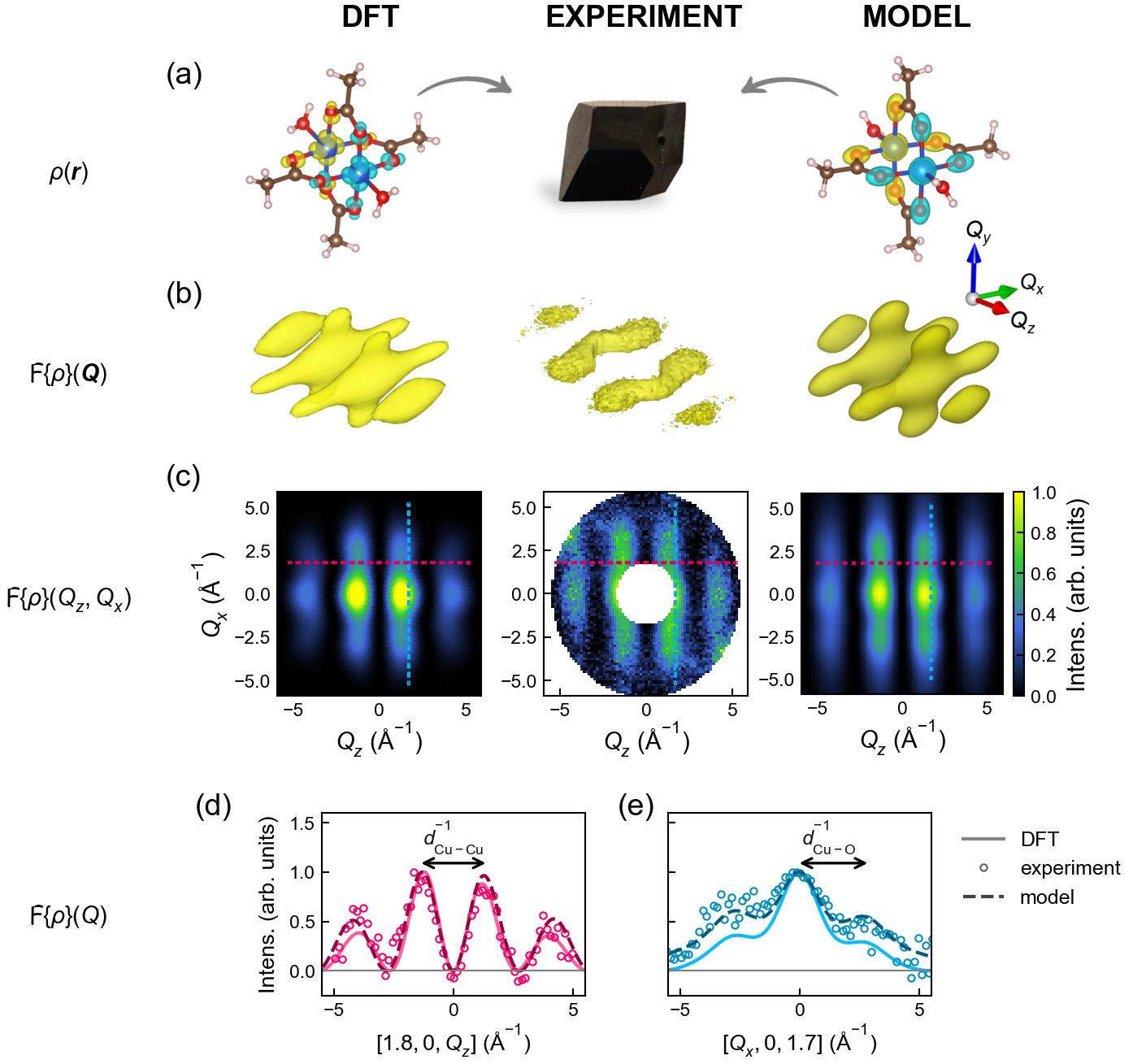}
\caption{Magnetic electron density and magnetic form factor of Cu(II) acetate monohydrate from DFT calculations, INS measurements, and analytical parametrization. (a) Broken-symmetry spin density, $\rho$, of the magnetic orbitals. (b-e) The magnetic form factor in reciprocal space (a Fourier transform of $\rho$) represented (b) as a  3D isosurface, (c) as a 2D cut in the equatorial ${Q_z}$--${Q_x}$ plane ($ac$ plane), and (d-e) as 1D cuts along the lines marked in panel (c). Note that part of the out-of-plane data is masked to enable comparison with the DFT results and analytical parametrization, which here are computed for a single dimeric unit and therefore do not include the interference effects arising from the two symmetry-inequivalent dimers present in the experiment (see \ref{app:outofplane}).
The inset in (d) indicates the reciprocal of the Cu-Cu distance projection onto the $xz$ plane.}
\label{fig:Fig2}
\end{figure*}

\begin{figure*}[t]
\centering
\includegraphics[width=1.0\linewidth]{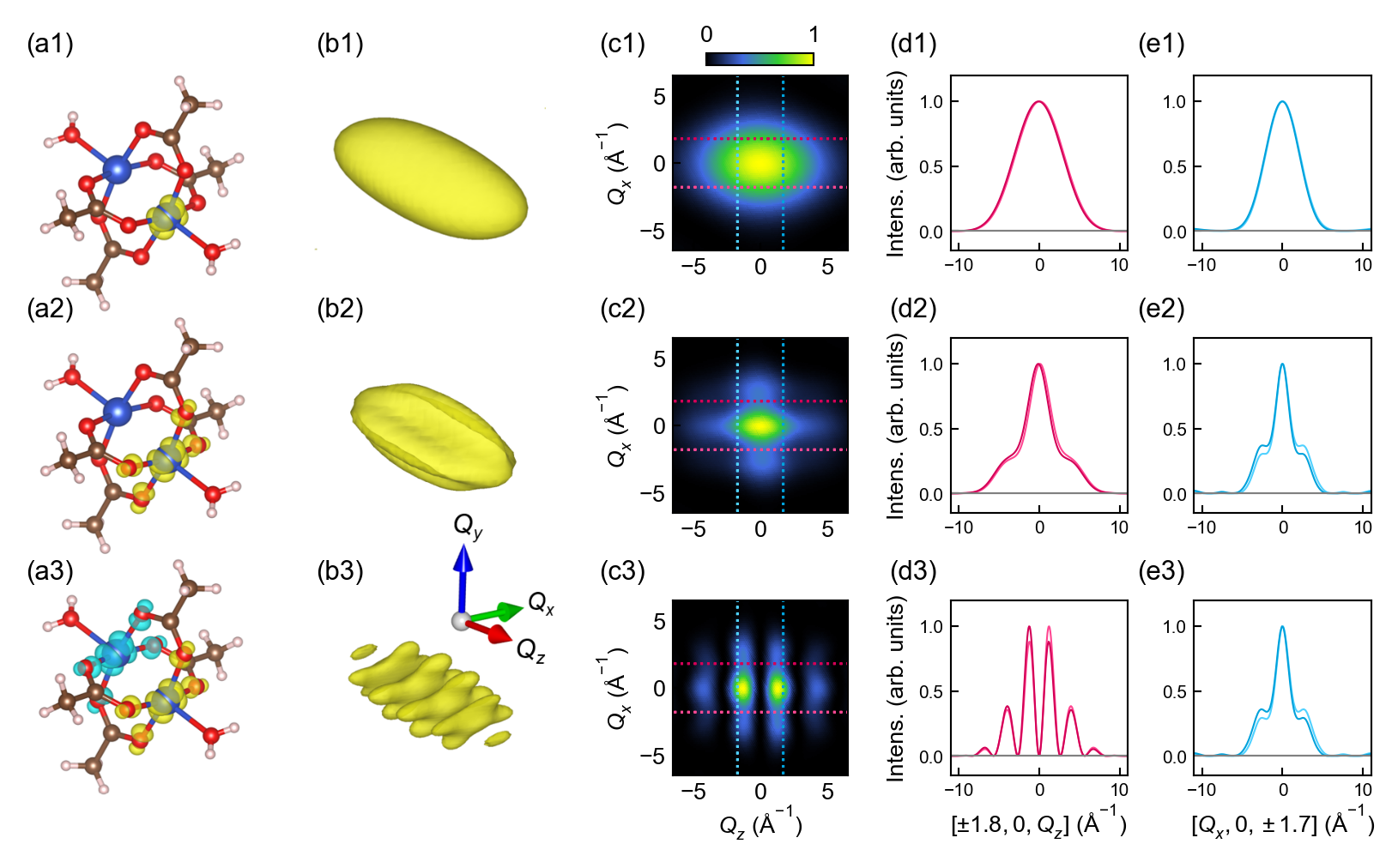}
\caption{Magnetic form factor from DFT and the impact of ligands. (a) The \spindensity{}, $\rho$, and the resulting form factor in (b) 3D, (c) 2D, and (d-e) 1D representations for (a1-e1) the Cu $d$ orbital alone with ligand density filtered out, (a2-e2) the full \spindensity{} of a single magnetic orbital, and (a3-e3) the complete broken-symmetry DFT spin density of the dimer. The comparison of the dark and light-colored cuts exemplifies the mirror asymmetry yet the inversion symmetry of Cu(II) acetate monohydrate. These panels realize the schematic program of Fig.~\ref{fig:intro} for the specific case of Cu(II) acetate monohydrate.}
\label{fig:Fig4}
\end{figure*}

The magnetic form factor encodes the spatial distribution of the \spindensity{} in reciprocal space and provides a direct link between neutron scattering measurements, analytical parametrization, and electronic structure calculations. In this section, INS data are combined with DFT and an analytical parametrization of the \spindensity{} to obtain a consistent and quantitative picture of the magnetic orbitals in Cu(II) acetate monohydrate, with particular emphasis on the role of ligand contributions.

For symmetrical dimers with $S_1 = S_2$, the $S^{zz}$ component of the diagonal cross-section of the singlet ($S=0$) to triplet ($S=1$) transition is given by
\begin{equation}
    \label{eq:Sz}
    S^{zz} (\mathbf{Q},\omega) = |E_{10}(\mathbf{Q})|^2 \delta(\hbar \omega - (E_1 - E_{0})),
\end{equation}
where the scattering amplitude involves the matrix element $E_{10} = \bra{\psi_{1}(\mathbf{r_1}, \mathbf{r_2})}e^{i\mathbf{Q}\cdot\mathbf{r_1}}\ket{\psi_{0}(\mathbf{r_1}, \mathbf{r_2})} = -\bra{\psi_{1}(\mathbf{r_1}, \mathbf{r_2})}e^{i\mathbf{Q}\cdot\mathbf{r_2}}\ket{\psi_{0}(\mathbf{r_1}, \mathbf{r_2})}$ and $\psi_{0}$ ($\psi_{1}$) denotes the singlet (triplet) spatial parts of the wavefunction, respectively. The energy $E_1 -E_0$ corresponds to the singlet–triplet gap.
Within the form factor approximation this becomes \cite{furrer_interference_1977,gudel_neutron_1977,furrer_magnetic_2013,guedel_direct_1979}
\begin{equation}
    \label{eq:Szz_formfactor}
    S^{zz} (\mathbf{Q},\omega) = \frac{1}{2} |F(\mathbf{Q})|^2(1-\mathrm{cos}(\mathbf{Q}\cdot \mathbf{R}))\delta (\hbar \omega - (E_1 - E_{0})),
\end{equation}
with $F(\mathbf{Q})$ the magnetic form factor, defined as the Fourier transform of the real-space \spindensity{} $\rho(\mathbf{r})$. The vector $\mathbf{R}$ indicates the dimer separation. 

The INS data are shown in the central column of Fig.~\ref{fig:Fig2}. Because neutrons probe spin--spin correlations in reciprocal space, the real-space \spindensity{} is not accessible directly; instead, the neutron structure factor is obtained by integrating the scattering intensity over an excitation-energy window centered around $\qty{37}{meV}$. The resulting three-dimensional isosurface and its projection onto the equatorial $(Q_z, Q_x)$--plane, shown respectively in Figs.~\ref{fig:Fig2}(b) and (c), reveal a pronounced oscillatory intensity pattern, reflecting the $(1-\mathrm{cos}(\mathbf{Q}\cdot \mathbf{R}))$ interference term of Eq.~\eqref{eq:Szz_formfactor}. The intensity modulation, of which a representative cut is shown in Fig.~\ref{fig:Fig2}(d), is aligned with $Q_z$, i.e., the reciprocal-space direction corresponding to the dimer axis (see Fig.~\ref{fig:dimer}). The measured periodicity is \qty{2.75}{{\AA}^{-1}}. Using the unit-cell parameters, the projection of the Cu--Cu distance onto the $ac$ plane is $d_{\mathrm{Cu-Cu}}^{ac}=\qty{2.19}{\AA}$, yielding the expected oscillation period
\begin{equation}
  \frac{2\pi}{d_{\mathrm{Cu-Cu}}^{ac}}=\qty{2.87}{{\AA}^{-1}},
\end{equation}
in good agreement with our measurements. A slight reduction of the observed periodicity is expected due to the decay of the magnetic form factor, which shifts intensity maxima towards the $\Gamma$ point. Note that the intensity oscillations at large $\boldsymbol{Q}$ should not be mistaken for \spindensity{} at large $r$ (e.g., on the \ce{H2O} ligand). The \spindensity{} is located around the Cu and O atoms only.

While the oscillatory dimer structure factor dominates the intensity along the Cu--Cu direction, it does not modulate the scattering perpendicular to the dimer axis, that is along $Q_x$. This allows the magnetic form factor along this direction to be extracted directly, as shown in Fig.~\ref{fig:Fig2}(e). The corresponding intensity profile exhibits a central peak accompanied by two satellite peaks at approximately $\pm\qty{3}{{\AA}^{-1}}$, with shallow dips in between. The asymmetry of the profile arises because the cut does not pass through the inversion center at the $\Gamma$ point. The presence of the satellite peaks indicates a partial delocalization of the Cu \spindensity{} onto the ligand O atoms. Considering the geometry of a single \ce{CuO4} plaquette, only two Cu--O bonds have a significant projection along $Q_x$, with an effective distance $d_{\mathrm{Cu-O}}=\qty{1.97}{\AA}$. This corresponds to a characteristic reciprocal-space scale
\begin{equation}
  \frac{2\pi}{d_{\mathrm{Cu-O}}}=\qty{3.19}{{\AA}^{-1}},
\end{equation}
consistent with the observed satellite positions and supporting the presence of a finite ligand \spindensity{}.

These experimental findings are corroborated both by analytical parametrization and DFT+$U$ calculations. The analytical parametrization of the 3D \spindensity{} introduced in Sec.~\ref{Analytical model} yields a ligand spin fraction of
$\overline{\rho}_{\mathrm{O}}/(\overline{\rho}_{\mathrm{Cu}}+4\,\overline{\rho}_{\mathrm{O}})\simeq\qty{5.3}{\%}$. The total \spindensity{} is depicted on top of the dimer unit in the right panel of Fig.~\ref{fig:Fig2} (a). Based on the best-fit values in Tab.~\ref{table:table1}, the Cu contribution to the \spindensity{} is described well by an oblate spheroidal distribution, with a larger extent within the \ce{CuO4} plane than along the dimer axis ($\sigma_{\mathrm{Cu}\parallel} > \sigma_{\mathrm{Cu}\perp}$). This is consistent with the $3d_{x^2-y^2}$ character of the $3d^9$ electron configuration in Cu(II) acetate monohydrate. In contrast, the O contributions to the \spindensity{} are elongated along the Cu--O bonds ($\sigma_{\mathrm{O}\parallel} > \sigma_{\mathrm{O}\perp}$), reflecting their dominant $2p$ character. The resulting isosurfaces and reciprocal-space projections, shown in Figs.~\ref{fig:Fig2}(b,c), reproduce the experimental features both qualitatively and quantitatively. The cuts in Figs.~\ref{fig:Fig2}(d,e) show how the fitted parametrization matches the positions of the maxima and minima, the overall decay with increasing $|\mathbf{Q}|$, and the satellite peaks, although the asymmetry is underestimated slightly. This effect is presumably a result of the Gaussian approximation and the assumption of a planar Cu environment, despite the Cu–O bonds being slightly tilted towards the center of the dimer unit.
A more extended comparison between the experiment and the analytical parametrization is shown in Figs.~S3, S4 and S5 in the Supplemental Material~\cite{supplemental}.

Independent confirmation is provided by the broken-symmetry DFT+$U$ calculations, which expose the magnetic orbitals of the singlet state as shown in the left panel of Fig.~\ref{fig:Fig2}(a). The magnetic structure factor in Figs.~\ref{fig:Fig2}(b--e) is obtained directly as a Fourier transform of this real-space \spindensity{}. The dominant Cu--Cu interference term, with a period of $d_\mathrm{Cu-Cu}^{-1}\simeq\qty{2.63}{{\AA}^{-1}}$, matches the experimental modulation. Cuts perpendicular to the dimer axis, containing the information about the magnetic form factor of the \ce{CuO4} unit, reproduce the satellite peaks and their asymmetry, confirming that these features originate from ligand contributions. The asymmetry stems from the absence of an inversion center on the Cu sites themselves. Integrating the DFT \spindensity{} within spheres centered on the atomic sites yields a ligand contribution of
$\overline{\rho}_{\mathrm{O}}/(\overline{\rho}_{\mathrm{Cu}}+4\,\overline{\rho}_{\mathrm{O}}) = 8.0 \%$,
which is in reasonable agreement with the value extracted from the analytical fit to the experimental data (Tab.~\ref{table:table1}). 

The systematic impact on the magnetic form factor of delocalization of \spindensity{} to the ligand is illustrated further in Fig.~\ref{fig:Fig4}, where successive inclusion of ligand contributions in the DFT spin density demonstrates their role in generating the observed modulation and satellite peaks. The cuts in the Cu-O plane (dark blue lines) in Figs.~\ref{fig:Fig4}(e1,e2) should be compared side-by-side with Figs.~\ref{fig:intro}(f,l).

In summary, the INS experiment, analytical parametrization, and DFT calculations consistently demonstrate that the magnetic structure factor of Cu(II) acetate monohydrate is governed by strong Cu--Cu interference resulting in the oscillating pattern along $Q_z$, while the magnetic form factor is altered by an appreciable \spindensity{} on the ligand O atoms. The combined analysis provides a quantitative estimate of the ligand contribution and demonstrates that subtle features of the reciprocal-space intensity reflect directly the real-space distribution of the magnetic orbitals, as expected from the analysis of the exemplary wavefunctions in Fig.~\ref{fig:intro}. The close agreement between theory and experiment shows that quantitative information on the \spindensity{} distribution can be extracted from the INS intensities of a magnetic excitation.

\section{Magnetic orbitals and interactions}
\label{sec:excitation}

\setlength{\tabcolsep}{8pt}
\renewcommand{\arraystretch}{1.5}
\begin{table*}[t]
\caption{Spectrum of the extended Hubbard dimer model Eq.~\eqref{eq:Hubbard_dimer_Hamiltonian_full} assuming $W_{ii}=4.94$~eV, $W_{ij}=2.45$~eV, $t=0.24$~eV and $J_\text{dir} = 0.024$~eV. The double-occupancy admixture $\delta \approx  \frac{2t}{W_{ii}-W_{ij}}= 0.193 $,
or precisely $\delta = \frac{1}{4t} \left( \sqrt{(W_{ii} - W_{ij} - J_\text{dir})^2 + (4t)^2} - (W_{ii} - W_{ij} - J_\text{dir})\right) = 0.188 $.
The double occupancy state with parity $P=+1$ is a singlet which mixes with the singly occupied singlet. The double occupancy state with parity $P=-1$, on the other hand, does not mix with the single occupied states and has an exact energy of $W_{ii}$.}
\centering
\begin{tabular}{cccc}
& \textbf{eigenstate} & \textbf{energy} & \textbf{energy (eV)} \\ \hline

\multirow{2}{*}{d. occ. $P=+1$} & $ \frac{1}{\sqrt{2}}\left(\ket{\uparrow\downarrow,0} - \ket{0,\uparrow\downarrow}\right)$ & $\frac{1}{2} \left( W_{ii} + W_{ij} + J_\text{dir} + \sqrt{(W_{ii} - W_{ij} - J_\text{dir})^2 + (4t)^2} \right) $ & 5.030 \\
 & $\textcolor{gray}{\;\;\;\;\;-\frac{\delta}{\sqrt{2}} \left(\ket{\uparrow,\downarrow} - \ket{\downarrow,\uparrow}\right)  }$ & $\approx  W_{ii} + \frac{4t^2}{W_{ii}-W_{ij}}$ & $\approx 5.032$ \\

 \hline
 d. occ. $P=-1$ & $\frac{1}{\sqrt{2}}\left(\ket{\uparrow\downarrow,0} + \ket{0,\uparrow\downarrow}\right)$ & $W_{ii}$ & 4.940
 \\ \hhline{====}
\multirow{3}{*}{triplet}  & $\ket{\uparrow,\uparrow}$ & \multirow{3}{*}{$W_{ij}-J_\text{dir}$} & \multirow{3}{*}{2.426} \\ 
\cline{2-2}
& $\frac{1}{\sqrt{2}}\left(\ket{\uparrow,\downarrow} + \ket{\downarrow,\uparrow}\right)$ &                               &                               \\ \cline{2-2}
& $\ket{\downarrow,\downarrow}$ &  &  \\ \hline
 \multirow{2}{*}{singlet} & $\frac{1}{\sqrt{2}}\left(\ket{\uparrow,\downarrow} - \ket{\downarrow,\uparrow}\right)$ & $\frac{1}{2} \left( W_{ii} + W_{ij} + J_\text{dir} - \sqrt{(W_{ii} - W_{ij} - J_\text{dir})^2 + (4t)^2} \right)$ & 2.384 \\
& $\textcolor{gray}{\;\;\;\;\;\;\;\;\;+\frac{\delta}{\sqrt{2}} \left( \ket{\uparrow\downarrow,0} - \ket{0,\uparrow\downarrow}\right)}$ & $\approx W_{ij} - \frac{4t^2}{W_{ii}-W_{ij}} + J_\text{dir}$ & $\approx 2.381$ \\ \hline
\end{tabular}
\label{tab:spectrum_of_H_dim}
\end{table*}

We can further represent the magnetic orbitals as maximally localized Wannier functions~\cite{marzari_maximally_1997}, capturing their full spatial wavefunction instead of only their density. In this section, we construct these Wannier functions and show how the interactions between them lead to a singlet ground state and the singlet-triplet splitting observed experimentally.

In this way we describe the low-energy physics of the system in terms of an effective Hubbard dimer model (HDM) in the basis of Wannier functions parametrized by cRPA~\cite{mazurenko_wannier_2007,miyake_screened_2008}.
This is motivated by the fact that, in the crystal structure, the $\mathrm{Cu}\text{-}\mathrm{Cu}$ dimer unit is isolated structurally from other dimers and the electronic structure near the Fermi level is characterized primarily by two molecular orbitals, with dominant $\mathrm{Cu}$ $d_{x^2-y^2}$ character, at a relatively short separation of $2.62$~$\mathrm{\AA}$. The many-body effects due to the strong interactions between electrons inhabiting these orbitals are therefore expected to be significant. Similar physics has been observed recently in other bulk materials, such as $\mathrm{VO}_{2}$ \cite{Haas2024_VO2} and the triniobium octahalide family $\mathrm{Nb}_{3}\mathrm{Cl}_{8}$ \cite{Aretz2025_Nb3X8, Grytsiuk2024_Nb3Cl8, carta_mlkvik_nb3cl8}.

It is therefore instructive to write down a model Hamiltonian for the Cu-Cu dimer unit as
\begin{equation}
\begin{aligned}
H_\text{HDM} = &-t \sum_{\sigma \in \{ \uparrow, \downarrow\}} (c_{i\sigma}^\dagger c_{j\sigma} + c_{j\sigma}^\dagger c_{i\sigma}) \\ &+ W_{ii} ( n_{i\uparrow}n_{i\downarrow} + n_{j\uparrow}n_{j\downarrow}) \\
&+ W_{ij} \, n_i n_j \\
& - J_\text{dir} \sum_{\sigma\sigma'} c_{i\sigma}^\dagger c_{j\sigma'}^\dagger c_{i\sigma'} c_{j\sigma},
\label{eq:H_dim_second_quant}
\end{aligned}
\end{equation}
where $i$ and $j$ label the two different Cu $d_{x^2-y^2}$ orbitals, which are connected by a hopping integral, $t$. The electrons in the dimer interact through the four-point integrals
\begin{align}
    W_{ijkl} &\equiv \langle ij| W | kl \rangle \\
    &= \iint w_i^*(\bm{r})w_k(\bm{r})  W(\bm{r}, \bm{r'}) w_j^*(\bm{r'}) w_l(\bm{r'})   \, d\bm{r} \, d\bm{r'},
    \label{eq:four_index}
\end{align}
which represent the effective screened Coulomb and exchange interactions of the electrons in the dimer orbitals $w_m$.
We consider the on-site Hubbard interactions, $W_{ii} \equiv \langle ii| W | ii \rangle$, the inter-site Hubbard interactions $W_{ij} \equiv \langle ij| W | ij \rangle$, and the direct exchange, $J_\text{dir} \equiv \langle ij| W | ji \rangle$~\cite{nakamura_respack_2021}.

\begin{widetext}
The dimer Hamiltonian is expressed in matrix form as
\begin{equation}
H_\mathrm{dim} = 
\bordermatrix{   ~ & \ketdeco{\uparrow,\downarrow} & \ketdeco{\downarrow,\uparrow} & \ketdeco{\uparrow\downarrow,0} & \ketdeco{0,\uparrow\downarrow} & \ketdeco{\uparrow,\uparrow} & \ketdeco{\downarrow,\downarrow}\cr
                \bradeco{\uparrow,\downarrow} & W_{ij} & -J_{\text{dir}} & -t & t & 0 & 0 \cr
                \bradeco{\downarrow,\uparrow} & -J_\text{dir} & W_{ij} & t & -t & 0 & 0 \cr
                \bradeco{\uparrow\downarrow,0} & -t & t & W_{ii} & 0 & 0 & 0 \cr
                \bradeco{0,\uparrow\downarrow} & t & -t & 0 & W_{ii} & 0 & 0 \cr
                \bradeco{\uparrow,\uparrow} & 0 & 0 & 0 & 0 & W_{ij}-J_\text{dir} & 0  \cr
                \bradeco{\downarrow,\downarrow} & 0 & 0 & 0 & 0 & 0 & W_{ij}-J_\text{dir}  \cr
                } \,,
\label{eq:Hubbard_dimer_Hamiltonian_full}
\end{equation}
\end{widetext}
where we have chosen the two-electron site-occupation basis $\left\{  \ket{\uparrow,\downarrow}, \ket{\downarrow,\uparrow}, \ket{\uparrow\downarrow,0}, \ket{0,\downarrow\uparrow}, \ket{\uparrow,\uparrow},\ket{\downarrow,\downarrow} \right\}$, in which the first (second) arrow in the ket denotes the spin value on the Cu site $i$(j). The state vector $\ket{\uparrow\downarrow,0}$ therefore represents a double occupation of site $i$ with no electron on $j$. The Hamiltonian is block-diagonal, dividing into two subspaces $\left\{  \ket{\uparrow,\downarrow}, \ket{\downarrow,\uparrow}, \ket{\uparrow\downarrow,0}, \ket{0,\downarrow\uparrow}\right\}$ and $\left\{\ket{\uparrow,\uparrow},\ket{\downarrow,\downarrow} \right\}$.

The on-site Hubbard interaction, $W_{ii}$, penalizes the two doubly occupied states $\ket{\uparrow\downarrow,0}$ and $\ket{0,\uparrow\downarrow}$. The inter-site Hubbard term, $W_{ij}$, denotes the interaction energy from equal occupation of the two Cu sites. In the first block, the direct exchange $J_\text{dir}$ acts as a spin-flip term connecting $\ket{\uparrow,\downarrow}$ with $\ket{\downarrow,\uparrow}$.  $J_\text{dir}$ also contributes to the on-site energies of the ferromagnetic triplets $\ket{\uparrow,\uparrow}$ and $\ket{\downarrow,\downarrow}$.

By a direct diagonalization of Eq.~\eqref{eq:Hubbard_dimer_Hamiltonian_full} we obtain its energy spectrum given in Table~\ref{tab:spectrum_of_H_dim}.

The ground state of the Hubbard dimer is mainly of singlet character $\frac{1}{\sqrt{2}}(\ket{\uparrow,\downarrow} - \ket{\downarrow,\uparrow})$ with minor admixture of the symmetrical double occupancy state\footnote{The symmetrical double occupancy state acquires an equal admixture of the $(\ket{\uparrow,\downarrow} - \ket{\downarrow,\uparrow})/\sqrt{2}$ singlet and its energy increases, which is the usual band repulsion.}, and its total energy is decreased by $4t^2 / (W_{ii}-W_{ij})$. This term favors the antiferromagnetic configuration of the singlet, which manifests as an antiferromagnetic superexchange (kinetic exchange).

\begin{figure*}[t]
\includegraphics[width=\linewidth]{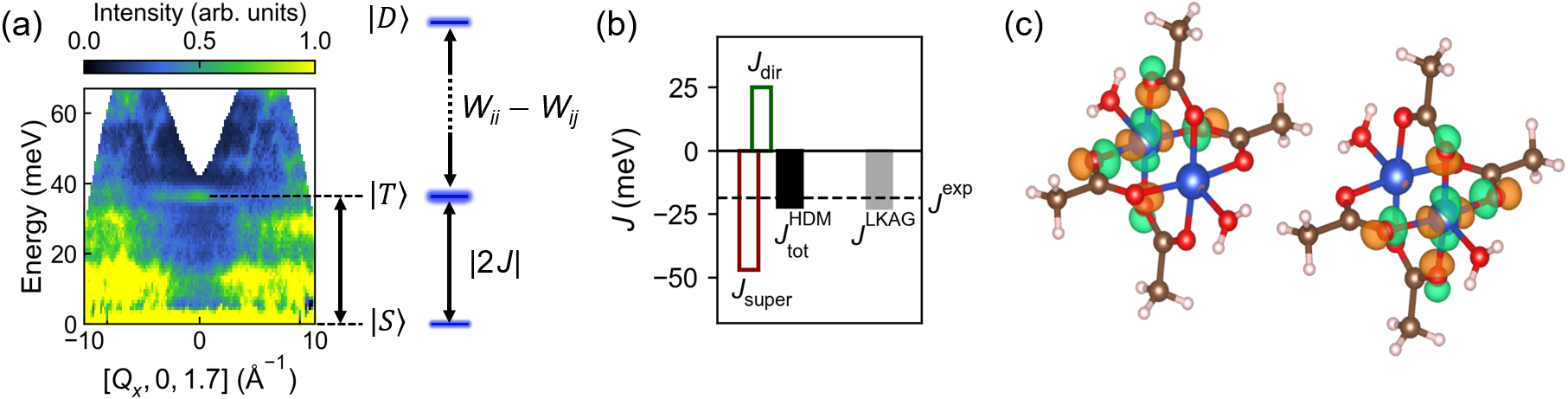}
\caption{Singlet-triplet splitting from the experiment and theory. (a) Energy-momentum neutron scattering plane. The momentum $Q$ corresponds to the light blue dashed line in Fig.~\ref{fig:Fig2}(c). The spectrum of the Hubbard dimer model (Eq.~\eqref{eq:Hubbard_dimer_Hamiltonian_full}) is displayed on the right, showing the singlet ($\ket{S}$), triplet ($\ket{T}$) and doubly-occupied states ($\ket{D}$) with their respective energy splittings. (b) Exchange interaction from the Hubbard dimer model, $J_\mathrm{tot}^\mathrm{HDM} = J_\mathrm{super} + J_\mathrm{dir} = \qty{-46.3}{meV} + \qty{24.0}{meV} = \qty{-22.3}{meV}$, compared with the result of the LKAG method, $J^\mathrm{LKAG} = -22.4$~meV, and the experimental value $J^\mathrm{exp} = -18.5$~meV. The singlet-triplet splitting $\Delta = |2 J|$ is then twice the exchange coupling in the convention of Eq.~\eqref{eq:Heisenberg_Hamiltonian_definition}. (c) Entangled magnetic orbitals in the form of their maximally localized Wannier functions.}
\label{fig:Fig5}
\end{figure*}

The triplet, on the other hand, is favored by the direct exchange, $J_\text{dir}$, because its antisymmetric spatial wavefunction minimizes the Coulomb repulsion. By contrast, the symmetric spatial wavefunction of the singlet renders its Coulomb repulsion higher.
One can show that the triplet energy is therefore lowered by approximately $2J_\text{dir}$ relative to the singlet.
Overall, the singlet-triplet splitting
\begin{align}
    \Delta &= E_\mathrm{T} - E_\mathrm{S} \\
    &\approx \frac{4t^2}{W_{ii}-W_{ij}}  - 2J_\text{dir}
\end{align}
takes the approximate form of the sum of the antiferromagnetic superexchange and the ferromagnetic direct exchange. We define the total exchange interaction consistent with the Heisenberg Hamiltonian of Eq.~\eqref{eq:Heisenberg_Hamiltonian_definition} as

\begin{equation}
    \Delta \equiv -2 J_\text{tot}^\mathrm{HDM} \,,
\end{equation}
resulting in the total exchange interaction

\begin{equation}
    J_\mathrm{tot}^\mathrm{HDM} \approx \underbrace{- \frac{2t^2}{W_{ii}-W_{ij}}}_{J_\text{super}}  + J_\text{dir} \,.
\label{eq:J_tot}
\end{equation}

To connect these model considerations with Cu(II) acetate monohydrate, we perform constrained random-phase approximation (cRPA) calculations. cRPA relies on the well defined energetic separation of the target states (in this case the Cu $d_{x^2-y^2}$ states) from the rest of the states in the crystal~\cite{Aryasetiawan2004}. The effective Coulomb interaction acting on the electron in the target states is then screened by all particle-hole excitations that occur outside these states~\cite{Aryasetiawan2004}. From these calculations we obtain the Hubbard dimer model parameters $W_{ii} = 4.94$~eV, $W_{ij}=2.45$~eV, and $J_\text{dir}=0.024$~eV, while the hopping integral $t=0.24$~eV can be inferred from the Wannierization procedure. 

The interdimer hopping is $t' = 0.018$~eV, introducing a negligible dispersion of the singlet-triplet excitation $J'/J\sim t'^2 / t^2 = 6 \times 10^{-3}$ relative to the singlet-triplet gap, consistent with no measurable dispersion of the singlet-triplet mode in INS. 

The fundamental excitation between the upper and lower Hubbard bands of the Hubbard dimer model $\approx W_{ii}-W_{ij} = 2.43$~eV agrees well with the band gap between the frontier bands in the DFT+$U$ calculation with $U=7.3$~eV, which is approximately $2.41$~eV (see the Supplemental Material~\cite{supplemental}).

From Eq.~\eqref{eq:J_tot} we obtain a total exchange coupling $J_\text{tot}^\mathrm{HDM} \approx J_\text{super} + J_\text{dir} = \qty{-46.3}{meV} + \qty{24.0}{meV} = -22.3$~meV, which is reasonably close to the experimental value of $J^\text{exp} =  -18.5$~meV, as shown in Figs.~\ref{fig:Fig5}(a-b). We confirm that a similar value $J^\mathrm{LKAG} = -22.4$~meV is obtained with the LKAG method. 
Also from the difference of the broken-symmetry and the high-spin (HS) state energies ~\cite{ruiz_broken_1999} $J^\mathrm{BS} = (E_{\mathrm{BS}}-E_{\mathrm{HS}})/2 = -25.5$~meV.

Note that the partitioning of the total exchange into superexchange and direct exchange depends on the choice of representation for the magnetic orbitals. In the basis of the canonical maximally localized WFs representing the natural magnetic orbitals the direct exchange is sizable, $\sim 50$\% of the antiferromagnetic superexchange.

We thus demonstrate that a Hubbard dimer model applied to Cu(II) acetate monohydrate captures the interactions between its magnetic orbitals quantitatively and can complement the calculations of their magnetic electron density based on broken-symmetry DFT. This further underlines the consistency between the presented experimental and theoretical methods.

\section{Discussion}
\label{sec:discussion}

We discuss our findings in three steps. First, we reflect on the results and then we discuss their generalization and applicability to other magnetic systems. Finally, we comment on their broader context and impact.

We have demonstrated that the \textbf{Q}-dependence of the INS intensities of a magnetic excitation reflects the real-space distribution of the magnetically active electrons, and thus contains microscopic information about the underlying spatial wavefunction envelope of the electrons. We presented a minimal parametrization to extract the delocalization of Cu magnetic electron density over ligand O atoms quantitatively and provided a description of the magnetic form factor within DFT. The construction of the maximally localized Wannier functions provides insight into the full spatial wavefunctions and reproduces the magnetic properties of Cu(II) acetate monohydrate. 

Beyond the isolated Heisenberg dimer Cu(II) acetate monohydrate, the approach is directly applicable to magnetic systems that host local or weakly dispersive excitations for which the momentum dependence of the INS intensity can be reliably modeled. The central requirement is that the dominant \textbf{Q}-dependence of the magnetic scattering cross section originates from a known, or minimally parametrizable, structure factor. Under this condition, the magnetic form factor -- and thus the real-space \spindensity{} -- can be extracted quantitatively. Any localized clusters of spins (dimers, trimers, and larger finite units) fall in principle  within this framework, including cases where weak inter-cluster interactions are present and can be treated perturbatively.

A first natural class of systems comprises all other isolated ferromagnetic or antiferromagnetic dimers, largely independent of their chemical composition or bridging ligands. Cu(II) acetate monohydrate represents a canonical example, but numerous related Cu-based molecular dimers have been studied and characterized extensively \cite{Cariati1982Magnetic,Sarkar2008CuII,Stylianou2008Synthesis,Shrivastava2025Synthesis}. Dimer physics is by no means restricted to Cu(II) ions: well-established examples include dimers based on Fe, Ni, Mn, Co, and Cr ions in both molecular and inorganic compounds \cite{Moseley2020Synthesis,Liu2017Tuning,Mennerich2006Antiferromagnetic,Zhang2024Unveiling,MAMERI20151370}. These systems provide clean realizations of localized magnetic excitations, for which the present methodology is ideally suited. If the center of inversion between the two metal ions in the dimer is broken, additional interactions such as the Dzyaloshinskii-Moriya interaction (DMI) are allowed \cite{moriya_anisotropic_1960,dzyaloshinsky_thermodynamic_1958,valenti_dzyaloshinskii-moriya_2000}. Extending our approach to asymmetric Heisenberg dimers with different amplitudes and sign of the DM vector promises insight into the way weakly ferromagnetic behavior in an antiferromagnet is reflected in the \spindensity{} distribution. This could potentially develop into a new route to capturing the DMI, which has been notoriously difficult to quantify experimentally \cite{Kuepferling_DMI_2023}.

A second broad class of relevant systems is that of single-molecule magnets (SMMs) \cite{Moreno_SMM_2021}. INS has long been employed to probe local spin excitations and crystal-field levels in single-ion magnets, including lanthanide complexes \cite{Vonci2017Magnetic,Dunstan2019Inelastic,Atkin2023Inelastic}, as well as in transition-metal molecular clusters such as \ce{Mn12} and \ce{Cr8} \cite{Hennion1997,Garlatti2019Unravelling}. In these systems, local magnetic excitations encode detailed information about the underlying spin Hamiltonian and magnetic anisotropy, making them particularly attractive targets for real-space spin-density reconstruction. For crystal-field excitations in single-ion magnets, the presented strategy is straightforward because the momentum dependence of the INS intensity is determined entirely by the magnetic form factor. In SMMs containing multiple magnetic ions, the finite size of the system allows excitation eigenstates and structure factors to be computed accurately using numerical approaches such as exact diagonalization. Alternatively, as demonstrated in Ref.~\cite{Garlatti2019Unravelling}, the measured neutron cross section can be fitted directly using a minimal parametrization, from which the magnetic form factor can be extracted. 

With appropriate care, the method can also be extended to strongly dimerized or frustrated quantum magnets that support predominantly local or nearly dispersionless modes. Examples of materials in which flat excitations have been observed include the Shastry–Sutherland system \ce{SrCu2(BO3)2}~\cite{SHASTRY19811069,Kageyama2000} and the fully frustrated magnet \ce{Ba2CoSi2O6Cl2}~\cite{Kurita_localized_2019}, the triangular lattice antiferromagnet ~\ce{BaLaCoTeO6} \cite{Park_anomalous_2024} and the square-kagome lattice antiferromagnet \ce{KCu6AlBiO4(SO4)5Cl}~\cite{Fujihala_squarekagome_2020}. In these cases, applicability depends on whether the structure factor in the localized-spin limit can be independently constrained or reliably modeled.

The application of an external magnetic field provides an additional probe of crystal-field excitations. By lifting degeneracies via the Zeeman interaction, the field renders excitation energies field-dependent and redistributes INS spectral weight, while leaving the magnetic form factor, and thus the real space \spindensity{}, to a good approximation unchanged. This enables the isolation of individual transitions and provides independent constraints on the \spindensity{}. This approach naturally reconnects to half-polarized neutron diffraction, where a field-polarized state allows reconstruction of the \spindensity{} from Bragg intensities. A similar strategy applies within the INS framework: in the limit of weakly interacting spins, transitions between Zeeman-split levels give rise to dispersionless excitations whose intensity is governed by the magnetic form factor, providing direct access to the underlying \spindensity{}.

Finally, we may consider systems with dispersive excitations for which the INS intensities also reflect the $\boldsymbol{Q}$-dependence of the magnetic excitation. Hence, dispersion-related contributions have to be separated before the magnetic form factor can be extracted. If the spin Hamiltonian is well established, the structure factor of the magnetic excitations can be calculated with well-established methods including linear spin-wave theory and the density matrix renormalization group (DMRG). For large dispersions, the integration of the experimental data over a large energy window poses a challenge to the extraction of reliable structure factors. Because instrumental resolution effects and background contributions depend on the energy transfer, larger systematic uncertainties could be introduced. Therefore, quantitative statements rely on the precise knowledge of the background and instrument parameters over the whole bandwidth of the magnetic excitation studied.

Although the primary focus of the presented study was to measure the magnetic form factor of a benchmark quantum magnet and to extract information about the microscopic electronic wavefunctions behind the macroscopic magnetic state, the impact of the approach extends further. A comprehensive understanding of the microscopic state is the prerequisite for engineering the electronic and magnetic properties in and out of equilibrium. DFT calculations are used extensively to explain out-of-equilibrium behavior \cite{fechner_magnetophononics_2018,giorgianni23}, underlining the importance of thorough benchmarking against experiments. This is especially important if the true magnetic state of the system cannot be captured by DFT, as is the case for quantum magnets. In the specific case of Cu(II) acetate monohydrate, our DFT calculations form the basis for predicting and understanding the non-equilibrium responses of the magnetic system to external stimuli, opening the path for nonequilibrium experiments investigating phenomena such as the modulation of the superexchange path by coherent phonon excitation or hydrostatic pressure.

\section{Conclusion}
\label{sec:conclusion}

On the textbook example of the singlet-to-triplet excitation of Cu(II) acetate monohydrate we demonstrated that the magnetic form factor can be extracted directly from the INS intensities of a magnetic excitation. We provided the proof of concept by extracting microscopic properties such as the 3D \spindensity{} distribution and the shape of the spatial electronic wavefunction on which it is based from the experimental data. 

Our state-of-the-art DFT calculations exhibit quantitative agreement with the experiment and provide insight into the microscopic electronic properties underlying the experimentally-observed magnetic form factor. This underscores the potential of broken-symmetry DFT calculations to determine spin-density distributions and magnetic form factors for explaining INS intensities in complex quantum magnets, thereby opening new perspectives on their exotic behavior. The Hubbard dimer model parametrized from DFT and cRPA calculations provides a simple yet complete view of the entangled magnetic orbital spectrum, correctly predicting the experimentally observed singlet-triplet splitting. Beyond INS and the textbook example of the isolated Heisenberg dimer, this benchmarking has a profound impact on experiments that rely heavily on DFT for their interpretation, first and foremost the study of magnetic exchange interactions out of equilibrium.

Our minimal analytical parametrization of the experimental data demonstrates that the \spindensity{} distribution can be reconstructed with quantitative precision from the $\boldsymbol{Q}-$dependence of the INS intensity of a magnetic excitation. Owing to its simplicity and low computational cost, this approach sets a gold standard for the quantification and treatment of the magnetic form factor in INS experiments. It can be applied easily to other magnetic systems with excitations whose structure factor can be  parametrized reliably, offering new insight into the microscopic parameters behind quantum magnetic phenomena. Magnetic systems of intense interest -- ranging from quantum magnets to altermagnets -- motivate addressing the challenge of dispersive excitations in the future, which includes obtaining high-resolution INS data with minimal background over large energy windows.

\begin{acknowledgments}
We acknowledge Mark de Vries for his contribution to the initial idea of this paper. We thank Nicola Colonna, Kristjan Eimre, Alfredo Fiorentino, Edward Baxter Linscott, Bruce Normand, Christian Rüegg, and Iurii Timrov for fruitful discussions. We acknowledge the Institut Laue Langevin (ILL), Grenoble, France for the provision of neutron beam time at PANTHER under Proposal No. 4-05-879 and the Materials and Life Science Experimental Facility (MLF) of the J-PARC, Tokai, Japan for the provision of neutron beam time at BL12 HRC under a user program (Proposal No. 2023B0236). This research was supported by the NCCR MARVEL, a National Centre of Competence in Research, funded by the Swiss National Science Foundation (grant number 205602). This research was supported by the Swiss National Science Foundation under grant 10001789. Computer time was provided by the Swiss National Supercomputing Centre (CSCS) under project No. lp18 and mr33.
\end{acknowledgments}

\section{Appendix} \label{Appendix}

\subsection{Roles of Cu--Cu, O--O, and Cu--O interferences}
\label{app:interference}

\begin{figure}
    \centering
    \includegraphics[width=1.0\linewidth]{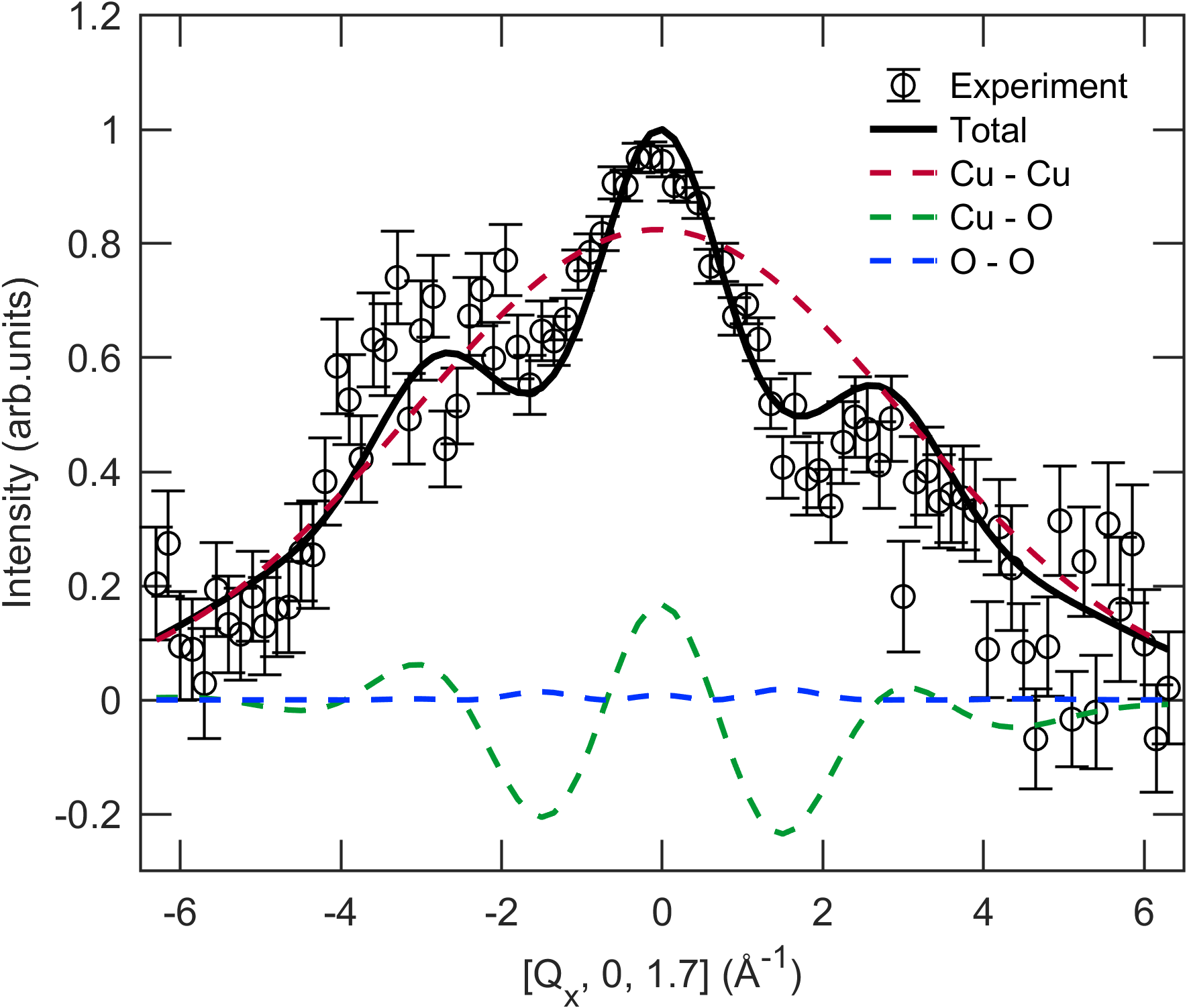}
    \caption{ Model-extrapolated contributions of the Cu-Cu, O-O and Cu-O interferences to the in-plane magnetic form factor shown in  Fig.~\ref{fig:Fig2}(e). The mixed term (green) plays a crucial role in shaping the momentum dependence.
}
    \label{fig:IP_FF_contributions}
\end{figure}

From the analytical parametrization, we can readily separate the contributions to the neutron scattering cross section arising from the pure Cu--Cu and O--O terms, as well as the mixed Cu--O terms. Indeed, the total neutron cross section can be seen as
\begin{equation}
S(\mathbf{Q}) = S_{\mathrm{Cu\!-\!Cu}}(\mathbf{Q}) 
     + S_{\mathrm{O\!-\!O}}(\mathbf{Q}) 
     + S_{\mathrm{Cu\!-\!O}}(\mathbf{Q}).
\end{equation}
It is instructive to analyze how these three terms shape the in-plane form  factor along $Q_x$ (Fig.~\ref{fig:Fig2}(e)).  A plot showing each individual contribution is provided in Fig.~\ref{fig:IP_FF_contributions}.
The Cu--Cu form factor produces only a broad, Gaussian-like curve. At the center, the Cu-Cu peak is enhanced and narrowed by $S_{\mathrm{Cu\!-\!O}}(\mathbf{Q})$, which generates both the dips at  $\pm\,\qty{1.75}{\text{\AA}^{-1}}$ and the two satellite peaks at $\pm\,\qty{3}{\text{\AA}^{-1}}$. 

The magnitude of each term is proportional to the corresponding localized  charge density. Specifically,
\[
S_{\mathrm{Cu\!-\!Cu}} \propto (\overline{\rho}_{\mathrm{Cu}})^{2}, \quad
S_{\mathrm{O\!-\!O}} \propto (\overline{\rho}_{\mathrm{O}})^{2}, \quad
S_{\mathrm{Cu\!-\!O}} \propto \overline{\rho}_{\mathrm{Cu}} \overline{\rho}_{\mathrm{O}}.
\]
This explains why the O--O contribution is essentially negligible, whereas the Cu--O term can significantly affect the total structure factor.

\subsection{Out-of-plane structure factor}
\label{app:outofplane}
Considering a simplified picture with point-like spin densities localized at the Cu sites, the magnetic neutron scattering for the singlet-to-triplet transition takes the form
\begin{equation}
    S (\mathbf{Q}) = \frac{3}{2} \left[1-\text{cos}\left(\mathbf{Q} \cdot \mathbf{d}_{12} \right)\right]+ \frac{3}{2} \left[ \left(1-\text{cos}(\mathbf{Q} \cdot \mathbf{d}_{34}\right)\right],
\label{eq:OOP CS}
\end{equation}
where $\mathbf{d}_{12}$ and $\mathbf{d}_{34}$ denote two dimers with distinct orientations within the unit cell.

\begin{figure}
\centering
\includegraphics[width=1\linewidth]{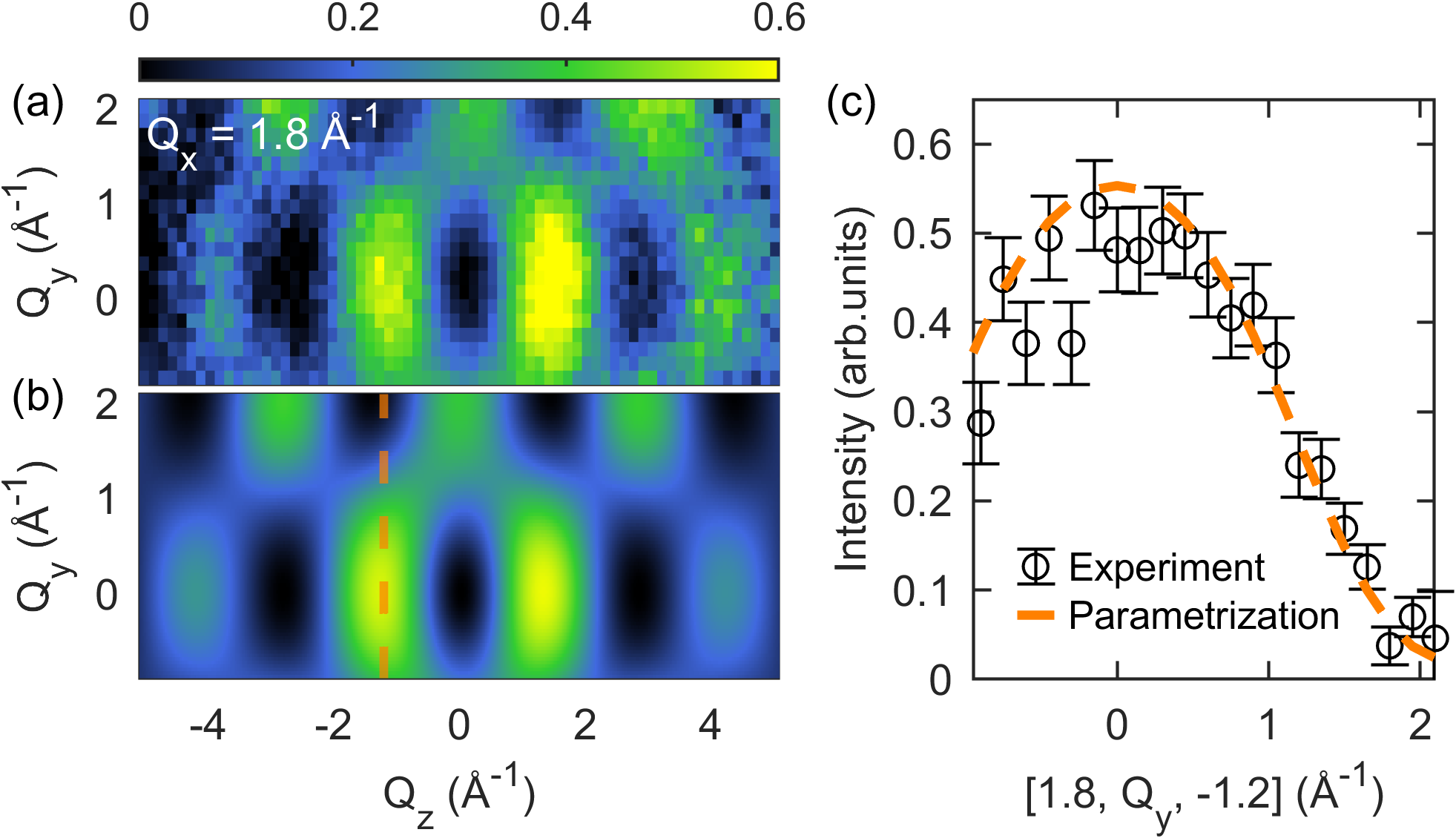}
\caption{(a) Neutron intensity in a plane perpendicular to the equatorial plane, with the dark pink dashed line in  Fig.~\ref{fig:Fig2}(c) as the horizontal axis. The checkerboard intensity pattern is the result of the interference between the two dimers in the unit cell. (b) The same plane simulated using the analytical parametrization. (c) Comparison between experiment and parametrization along the orange dashed line in panel (b).}
\label{fig:Fig7}
\end{figure}

In the horizontal scattering $(Q_z, Q_x)$--plane, the structure factor is modulated only along a single direction. This follows directly from Eq.~\eqref{eq:OOP CS} when the scattering vector is of the form $\mathbf{Q} =[Q_x, 0, Q_z]$:
\begin{equation}
    S (\mathbf{Q}) = 3 \left[1 -\text{cos}\left(Q_z\,d_{\mathrm{Cu-Cu}}^{ac}\right)\right].
\end{equation}
However, the inclusion of a finite $Q_y$ component in the scattering vector introduces an additional momentum-space modulation. This arises because the $y$-component of the dimer separation, $d_y$, changes sign between the two dimers. The structure factor then becomes
\begin{align}
S(\mathbf{Q}) &= \frac{3}{2}\left[1-\cos\left(Q_z\,d_{\mathrm{Cu-Cu}}^{ac}+Q_y d_y \right)\right] \notag \\[6pt]
&\quad + \frac{3}{2}\left[1-\cos\left(Q_z\,d_{\mathrm{Cu-Cu}}^{ac} - Q_y d_y\right)\right],
\end{align}
which can be compactly rewritten using trigonometric identities as
\begin{equation}
S(\mathbf{Q}) = 3 \left[1 - \cos\left(Q_z\,d_{\mathrm{Cu-Cu}}^{ac}\right)\cos\left(Q_y d_y\right)\right].
\end{equation}
This additional modulation along $Q_y$ is visible in the neutron-scattering data thanks to the vertical coverage of PANTHER detector. In Fig.~\ref{fig:Fig7}(a) we show the neutron intensity in a selected ($Q_z$, $Q_y$)--plane. The checkerboard pattern clearly reveals the double modulation. The analytical parametrization reproduces this behavior well, as demonstrated by the corresponding simulated plane in Fig.~\ref{fig:Fig7}(b). Finally, we show the direct comparison of a line cut in Fig.~\ref{fig:Fig7}(c). The reduction in intensity along the cut comes from the combined effect of the cosine modulation along $Q_y$ and the magnetic form factor.

\subsection{Singlet and broken-symmetry DFT} \label{app:singlet_and_DFT}

The singlet state

\begin{equation}
    |S \rangle = \frac{1}{\sqrt{2}} \left(\ket{\uparrow, \downarrow} - \ket{\downarrow ,\uparrow} \right)
\end{equation} is a superposition of two broken-symmetry (BS) states  
$\ket{\uparrow, \downarrow }$ and $\ket{\downarrow ,\uparrow}$, each of them being a single Slater determinant of the spin orbitals on the left and right Cu sites. For instance, $\ket{\downarrow, \uparrow} = |L \beta \,R\alpha| = \frac{1}{\sqrt{2}}\left( L(\bm{r}_1)\beta(\bm{r}_1)R(\bm{r}_2)\alpha(\bm{r}_2) + R(\bm{r}_1)\alpha(\bm{r}_1)L(\bm{r}_2)\beta(\bm{r}_2)\right)$ denotes a single Slater determinant with the left magnetically active orbital $L$ in minority spin $\beta$ times the right magnetically active orbital $R$ in majority spin $\alpha$. 

This means that the singlet spin density is zero everywhere, which does not hold for the broken-symmetry solutions individually~\cite{trushin_avoiding_2023}.\footnote{The alternative view of this freezing is that of collective excitations in a solid (a charge-density wave or a spin-density wave) softening under a varying external potential with their frequency tending to zero until they appear as a static symmetry-broken wave~\cite{perdew_interpretations_2021}.} Such spin contamination leads to an incorrect nonzero spin density in space, even though the integrated spin density is still zero. This breaks the symmetry of the physical Hamiltonian.

Indeed, in open-shell systems, the Kohn-Sham (KS) Hamiltonian and the KS determinant (the resulting wavefunction) both present a lower symmetry than the physical Hamiltonian and the physical wavefunction~\cite{trushin_avoiding_2023}. The KS formalism is still \textit{formally} correct, but the symmetry lowering alters the partitioning of the total energy into a part obtained by the KS determinant and the correlation energy (the remainder). This can lead to errors due to the approximations in \textit{practical} DFT calculations~\cite{trushin_avoiding_2023}. Note that in the hypothetical case of an \textit{exact} DFT calculation, the symmetry breaking would not occur~\cite{trushin_avoiding_2023}. In practical calculations, however, approximations are always made.

On the other hand, the BS solutions provide a substantial insight, because they expose in a static way the states between which the true ground state fluctuates~\cite{neese_definition_2004,anderson_more_1972}.

Moreover, the BS calculation actually leads to a better total energy prediction than a DFT calculation where the spin density is forced to be zero everywhere (spin polarization is switched off)~\cite{trushin_avoiding_2023}. This is exemplified in the calculation of the exchange constant, which can be obtained from the broken-symmetry energy $E_{\mathrm{BS}}$ and a high-spin (HS) state energy $E_{\mathrm{HS}}$ as~\cite{ruiz_broken_1999}

\begin{equation}
    \label{eq:yamaguchi}
    2 J = \frac{E_{\mathrm{BS}}-E_{\mathrm{HS}}}{2 S_1 S_2 + S_2}
\end{equation}
which for Cu(II) acetate monohydrate with $S_1=S_2=1/2$ leads to  $J^\mathrm{BS} = (E_{\mathrm{BS}}-E_{\mathrm{HS}}) / 2 = -25.5$~meV, comparing to the experimental value $J^\text{exp} =  -18.5$~meV, while the non-spin-polarized (NS) calculation would lead to a completely incorrect prediction of $(E_{\mathrm{NS}} - E_{\mathrm{HS}})/2 = -330.5$~meV.

\begin{figure}
    \centering
    \includegraphics[width=0.80\linewidth]{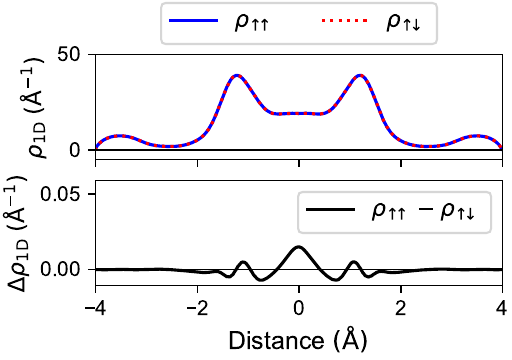}
    \caption{Top panel: The DFT+$U$-calculated charge density of the $\ket{\uparrow\uparrow}$ and $\ket{\uparrow\downarrow}$ states along the Cu-Cu direction (integrated in the plane perpendicular to that direction) and (bottom panel) their difference, illustrating Eq.~\eqref{eq:charge_density}.
}
    \label{fig:charge_density_difference}
\end{figure}

Finally, the electron \textit{density} resulting from a broken-symmetry calculation is correct if the overlap density between the Cu$_1$ and Cu$_2$ localized states is negligible. This can be shown as follows:  

We analyze how the charge density varies depending on the particle-exchange symmetry of the orbital wavefunction $\Psi(\mathbf{r_1}, \mathbf{r_2})$ of two interacting particles, which we assume can be written as:
\begin{equation}
    \Psi(\mathbf{r_1}, \mathbf{r_2})^\pm = \frac{1}{\sqrt 2} \left[ L(\mathbf{r_1})R(\mathbf{r_2}) \pm L(\mathbf{r_2})R(\mathbf{r_1}) \right],
\end{equation}
where $L$ and $R$ are real single particle orbitals centered around Cu$_1$ and Cu$_2$ sites, respectively.
$\Psi^+$ goes with the spin singlet state and $\Psi^-$ with the triplet state.  
The charge density for the first particle is:
\begin{equation}
\begin{split}
    \rho^1(\mathbf r) 
     = & e\int d\mathbf{r_1} d\mathbf{r_2}  \delta(\mathbf{r} - \mathbf{r_1}) |\Psi^\pm(\mathbf{r_1}, \mathbf{r_2})|^2 \\
     = & e\int d\mathbf{r_2}  |\Psi^\pm(\mathbf{r}, \mathbf{r_2})|^2 \\
     = & \frac{e}{2}|L(\mathbf{r})|^2 \int d\mathbf{r_2} |R(\mathbf{r_2})|^2
      + \frac{e}{2}|R(\mathbf{r})|^2 \int d\mathbf{r_2} |L(\mathbf{r_2})|^2\\
    & \pm e L(\mathbf{r})R(\mathbf{r}) \int d\mathbf{\mathbf{r_2}} R(\mathbf{r_2})L(\mathbf{r_2}) \\
    = & \frac{e}{2} \left(|L(\mathbf{r})|^2 A_a
      + |R(\mathbf{r})|^2 A_b \right)  \pm e L(\mathbf{r})R(\mathbf{r}) S_{LR} .\\
\end{split}
\end{equation}
Here, $S_{LR} = \int d\mathbf{r_2} R(\mathbf{r_2})L(\mathbf{r_2})$ is the overlap integral.
If $\phi$ is normalized $A_a=A_b=\int d\mathbf{r_2} |\phi(\mathbf{r_2})|^2  = 1 $, and

\begin{equation}
\label{eq:charge_density}
\begin{split}
    \rho^1(\mathbf r) 
     = & \frac{e}{2} \left(|L(\mathbf{r})|^2
      + |R(\mathbf{r})|^2 \right)  \pm e L(\mathbf{r})R(\mathbf{r}) S_{LR} .
\end{split}
\end{equation}\\

That is, the charge density at a point $\mathbf r$ is the average of the probability densities from the two orbitals $\pm$ the overlap term, which is only finite in the region of space where the two orbitals overlap and is weighted by the overlap integral $S_{LR}$.
The charge density for the second particle is the same, which is a property of the indistinguishability of the particles: $|\Psi^\pm(\mathbf{r_1}, \mathbf{r_2})|^2 = |\Psi^\pm(\mathbf{r_2}, \mathbf{r_1})|^2$.
This implies that the spin density $\rho_s = \rho^1 - \rho^2$ is always zero for both symmetrizations of the wavefunction.
Finally, if we consider that the orbital wavefunction is a broken-symmetry Hartree product $\Psi(\mathbf{r_1}, \mathbf{r_2}) = L(\mathbf{r_1^\uparrow})R(\mathbf{r_2^\downarrow}) $, the overlap term in Eq.~\eqref{eq:charge_density} vanishes, and a finite spin density is possible.

This is demonstrated numerically in Figure~\ref{fig:charge_density_difference}.

\FloatBarrier

\vspace{0.5cm}

\bibliography{Bibliography}

\clearpage
\onecolumngrid

\section*{Supplemental Material}

\renewcommand{\thefigure}{S\arabic{figure}}
\renewcommand{\thetable}{S\arabic{table}}
\renewcommand{\thesection}{S\arabic{section}}

\setcounter{figure}{0}
\setcounter{table}{0}
\setcounter{section}{0}

\renewcommand{\vec}[1]{\boldsymbol{#1}}

\section{Crystal growth}

Deuterated copper acetate hydrate \ce{Cu2(CD3COO).4(D2O)2} crystallizes in the monoclinic space group C2/c (no. 15) with lattice parameters $a=\qty{13.168}{\AA}$, $b=\qty{8.564}{\AA}$, $c=\qty{13.858}{\AA}$, and $\beta =\ang{117.02}$ \cite{de_meester_refined_1973}; it is also known as the mineral Hoganite. 
\ce{Cu2(CD3COO).4(D2O)2} was synthesized from copper(II) oxide (CuO, $\qty{99.995}{\%}$, Thermo Scientific), deuterated acetic acid (\ce{CD3COOD}, $\qty{99.5}{\%}$ D, Thermo Scientific), and deuterated water (\ce{D2O}, $\qty{99.8}{\%}$ D, Armar). CuO ($\qty{25}{\gram}$) was filled into a $\qty{250}{\milli\liter}$ flask equipped with an air-cooled condenser. Under argon atmosphere, acetic acid ($\qty{50}{\gram}$) and water ($\qty{50}{\gram}$) were added. The mixture was stirred and heated on an oil bath to $\qty{90}{\degreeCelsius}$ until the \ce{CuO} had dissolved completely. After about two days, the product was cooled to room temperature, filtered off, and recrystallized from \ce{D2O} for purification. 

For crystal growth, a saturated solution was prepared by dissolving the material in \ce{D2O} at $\qty{60}{\degreeCelsius}$. A few drops of \ce{CD3COOD} were added, just enough to avoid hydrolysis. The saturated solution was filtered and slowly cooled to room temperature in a closed flask. Small millimeter size crystals were selected and used as seed crystals in further crystallizations. Crystals of $\qty{3.5}{\gram}$ and $\qty{2}{\centi\meter}$ size could be obtained within one week. The best results were achieved in close-to-neutral solutions. Fully faceted crystals were obtained with 200, 002, 110, 110, and 111, 11-2 faces, in order of decreasing surface area. The specimen used in the experiments described in the main text is shown in Fig.~2(a). The crystal quality was verified by single-crystal X-ray diffraction, from which the parameters of the unit cell at ambient temperature were refined to $\text{a} = \qty{13.1722(2)}{\AA}$, $\text{b} = \qty{8.5615(1)}{\AA}$, $\text{c} = \qty{13.8610(2)}{\AA}$ and $\beta = \ang{117.026(2)}$, in close agreement with the literature values mentioned above.

\section{Preliminary neutron data}

\begin{figure}[h]
    \centering
    \includegraphics[width=0.75\linewidth]{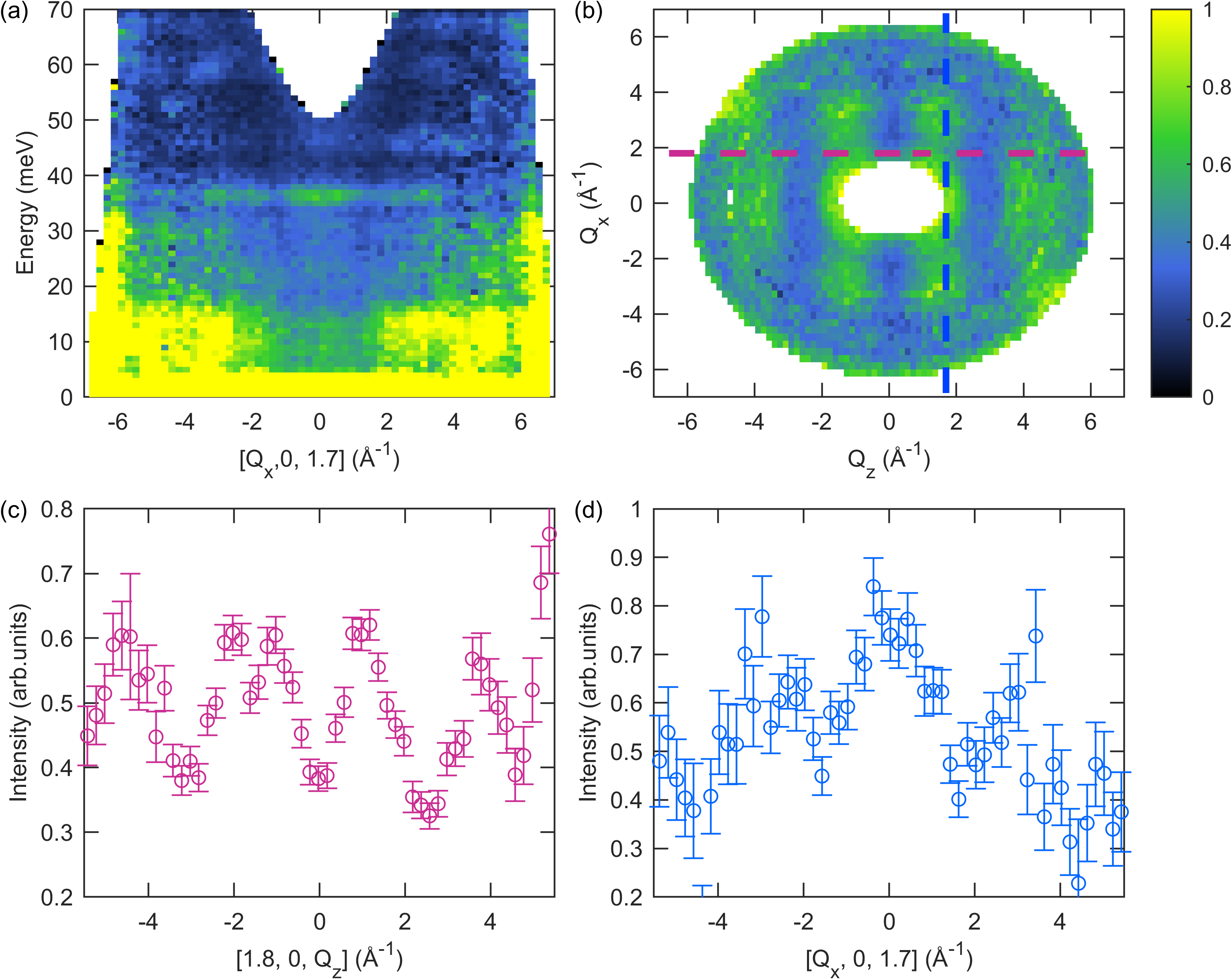}
    \caption{(a) $(Q_x$, $E$)--plane, where the $x$-axis corresponds to the blue dashed line of panel (b). The singlet-to-triplet excitation is visible at about \qty{37}{meV}. (b) Intensity in the equatorial $(Q_z, Q_x)$--plane obtained by integrating over an energy window centered at the excitation energy. (c) Line cut along the pink dashed line of panel (b) showing the cosine modulation arising from the dimer geometry. (d) In-plane form factor extrapolated along the blue dashed line of panel (b).
}
    \label{fig:hrc}
\end{figure}

Fig.~\ref{fig:hrc} provides an overview of neutron scattering data collected using the High Resolution Chopper Spectrometer (HRC) at the Materials and Life Science Experimental Facility (MLF), J-PARC~\cite{itoh_high_2011,itoh_progress_2019,ueta_sample_2024}. An incident neutron energy of \qty{100}{meV} was used, and the sample was cooled to 10~K. The ($Q_x$, $E$)--plane in Fig.~\ref{fig:hrc}(a), which shows the triplet excitation at approximately 37~meV, is directly comparable to Fig.~5(a) of the main text. Fig.~\ref{fig:hrc}(b) displays the momentum dependence of the magnetic mode in the equatorial $(Q_z, Q_x)$--plane, while Figs.~\ref{fig:hrc}(c, d) show line cuts along the purple and blue dashed lines of Fig.~\ref{fig:hrc}(b), respectively. The HRC data are in agreement with the corresponding measurements done on PANTHER (Fig.~2 of the main text).

\section{Analytical parametrization}
\subsection{Experimental background}
\begin{figure} 
    \centering 
    \includegraphics[width=0.75\linewidth]{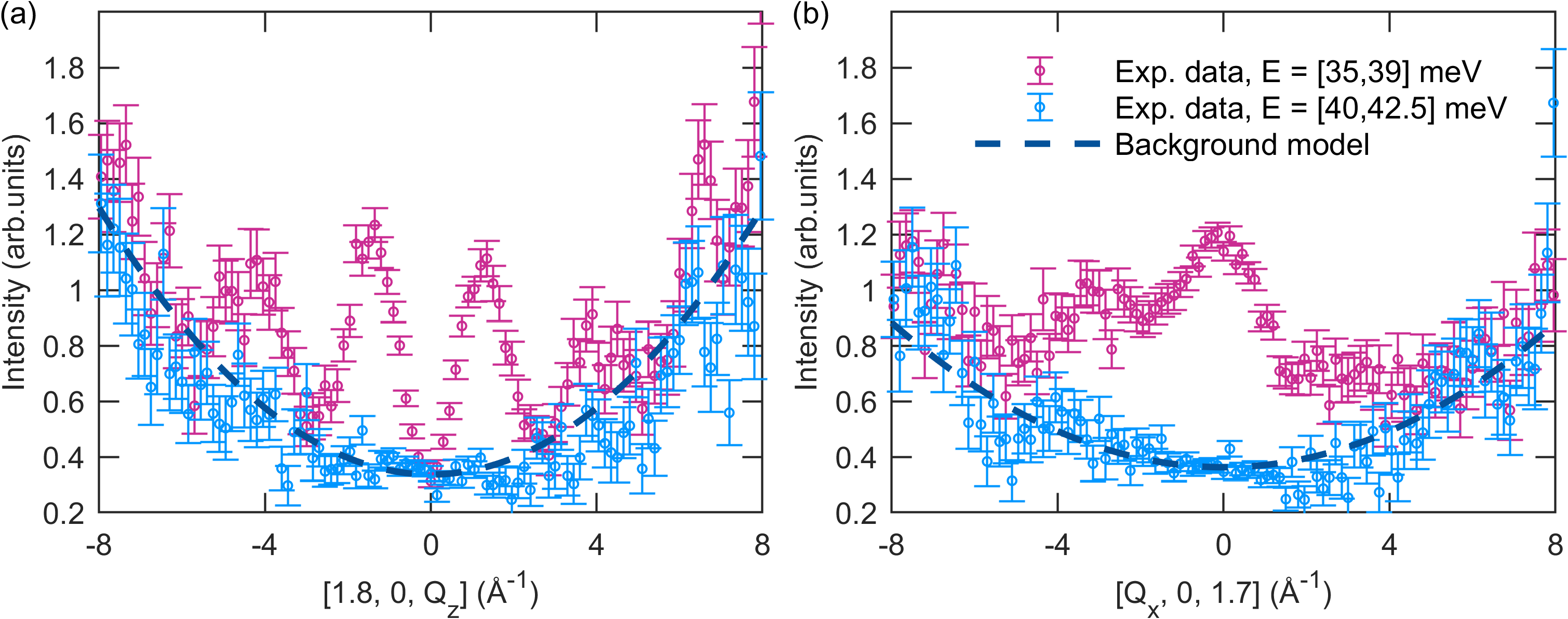}
\caption{
(a) Purple: the same line cut as in Fig.~3(d) of the main text prior to background subtraction. Blue: data used for background modeling, together with the best-fit model (dashed line).
(b) Purple: the same line cut as in Fig.~3(e) of the main text prior to background subtraction. Blue: data used for background modeling, together with the best-fit model (dashed line).
}
    \label{fig:bg_fit}
\end{figure}

To isolate the magnetic structure factor of the singlet-to-triplet excitation in the INS data collected on PANTHER, the phonon-dominated background was modeled using a function quadratic in the magnitude of the scattering vector $\mathbf{Q}$, consistent with the typical $|\mathbf{Q}|^2$ dependence of neutron structure factors for lattice vibrations
\begin{equation}
    \textrm{background}(\mathbf{Q})=a \, |\mathbf{Q}|^2 + b,
\end{equation}
where $a$ and $b$ are the free parameter of the model.
Neutron data were integrated in an energy window immediately above the singlet-to-triplet excitation, thus empty of any magnetic signal, and given as input to the fitting routine. Two representative line cuts are shown in Fig.~\ref{fig:bg_fit}. Subsequently, we subtracted the modeled background from the INS signal integrated in the excitation-energy window at about 37~ meV.

\subsection{Best-fit values as a function of the voxel side}
The analytical parametrization described in Sec.~II C was fitted to the experimental data after background subtraction. Each experimental intensity point corresponds to an integration over a reciprocal-space voxel of side $dQ$,
\begin{equation}
I^{(\mathrm{exp})}(\mathbf{Q}) =
\int_{-\frac{dQ}{2}}^{\frac{dQ}{2}} \! \int_{-\frac{dQ}{2}}^{\frac{dQ}{2}} \! \int_{-\frac{dQ}{2}}^{\frac{dQ}{2}}
I(\mathbf{Q} + \mathbf{q}) \,
\mathrm{d}q_x \, \mathrm{d}q_y \, \mathrm{d}q_z .
\end{equation}
To improve the accuracy of the fit, the model intensity was evaluated at multiple reciprocal-space points within each voxel and averaged. Table~\ref{tab:SM_fit_parameters_final} summarizes the best-fit values of the parameters for the different choices of $dQ$, which show only minor variations. We performed the fit employing two different masks for the experimental data, one being more strict in removing spurious scattering signal. In addition, we tested the inclusion of a flat background term as an extra parameter. The refined $ \frac{\overline{\rho}_{\mathrm{O}}}{\overline{\rho}_{\mathrm{Cu}}+4\,\overline{\rho}_{\mathrm{O}}}$ remains essentially unaffected under these variations, while the $\sigma$ parameters show only minor changes, preserving the inequalities $\sigma_{\mathrm{Cu}\parallel} > \sigma_{\mathrm{Cu}\perp}$ and $\sigma_{\mathrm{O}\parallel} > \sigma_{\mathrm{O}\perp}$.

\renewcommand{\arraystretch}{1.1}
\begin{table*}[t]
\caption{Best-fit values of the parameters as a function of $dQ$. Uncertainties correspond to one standard deviation.}
\begin{ruledtabular}
\begin{tabular}{c c c c c c c}
$dQ \, (\mathrm{\AA}^{-1})$ &
$A$ &
    $\displaystyle \frac{\overline{\rho}_{\mathrm{O}}}{\overline{\rho}_{\mathrm{Cu}}+4\,\overline{\rho}_{\mathrm{O}}} \, (\%)$  &
$\sigma_{\mathrm{Cu}\perp} \, (\mathrm{\AA}^{-1}) $ &
$\sigma_{\mathrm{Cu}\parallel} \, (\mathrm{\AA}^{-1})$ &
$\sigma_{\mathrm{O}\parallel}\, (\mathrm{\AA}^{-1})$ &
$\sigma_{\mathrm{O}\perp}\, (\mathrm{\AA}^{-1})$  \\ \hline
0.100 & 7.05(3)$\times10^{-5}$ & 5.31(4) & 0.196(1) & 0.237(1) & 0.403(5) & 0.286(5)  \\
0.125 & 6.91(3)$\times10^{-5}$ & 5.29(5) & 0.183(1) & 0.230(1) & 0.430(5) & 0.271(5) \\
0.150 & 6.91(4)$\times10^{-5}$ & 5.28(5) & 0.184(1) & 0.230(1) & 0.427(6) & 0.270(6)  \\
0.175 & 6.85(4)$\times10^{-5}$ & 5.28(5) & 0.178(1) & 0.226(1) & 0.440(7) & 0.268(7) \\
0.200 &  6.96(5)$\times10^{-5}$ & 5.33(8) & 0.176(1) & 0.225(1) & 0.447(8) & 0.259(8) \\
0.225 & 6.80(6)$\times10^{-5}$ & 5.20(8) & 0.176(1) & 0.225(1) & 0.441(9) & 0.255(9)  \\
0.250 & 6.66(6)$\times10^{-5}$ & 5.03(8) & 0.174(2) & 0.224(1) & 0.43(1) & 0.25(1)  \\
0.275 & 6.78(8)$\times10^{-5}$ & 5.2(1) & 0.175(2) & 0.224(1) & 0.44(1) & 0.26(1) \\
0.300 & 6.84(8)$\times10^{-5}$ & 5.3(1) & 0.178(2) & 0.223(1) & 0.46(1) & 0.28(1) \\
\end{tabular}
\end{ruledtabular}
\label{tab:SM_fit_parameters_final}
\end{table*}
\renewcommand{\arraystretch}{1.0}

\subsection{Extended comparison between the experiment and the model}
\begin{figure} 
    \centering 
    \includegraphics[width=1\linewidth]{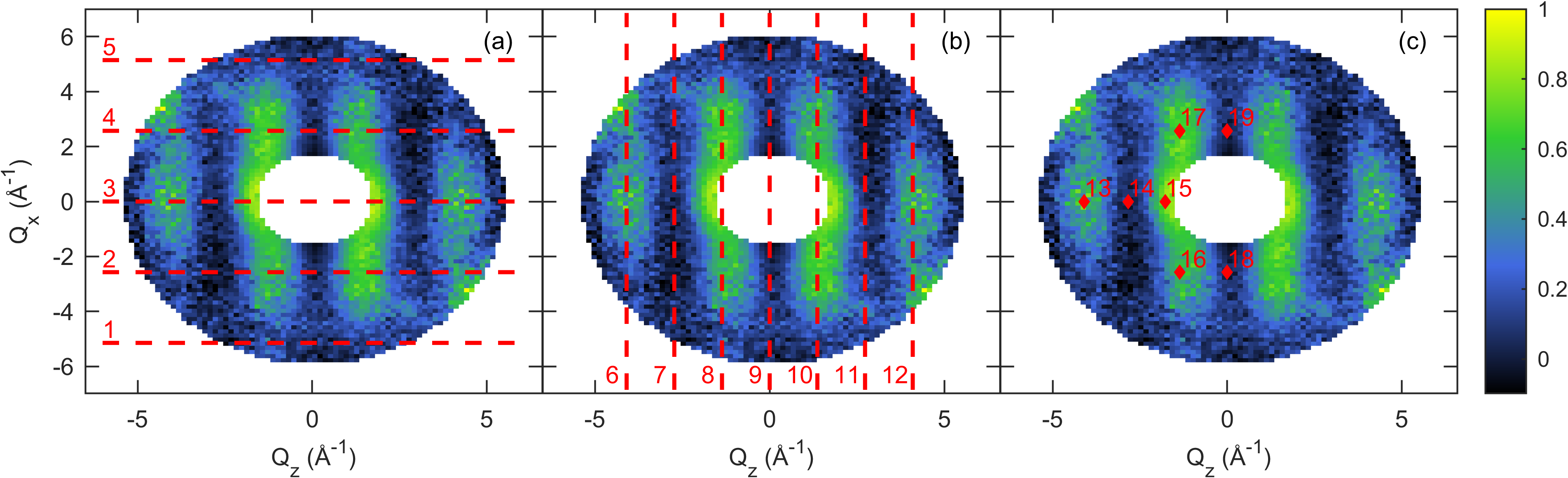}
    \caption{Three replicas of the excitation momentum dependence in the equatorial $(Q_z, Q_x)$--plane, together with a finite set of line cuts selected for comparison between the experiment and the analytical parametrization.
}
    \label{fig:many_cuts_maps}
\end{figure}

\begin{figure} 
    \centering
    \includegraphics[width=1\linewidth]{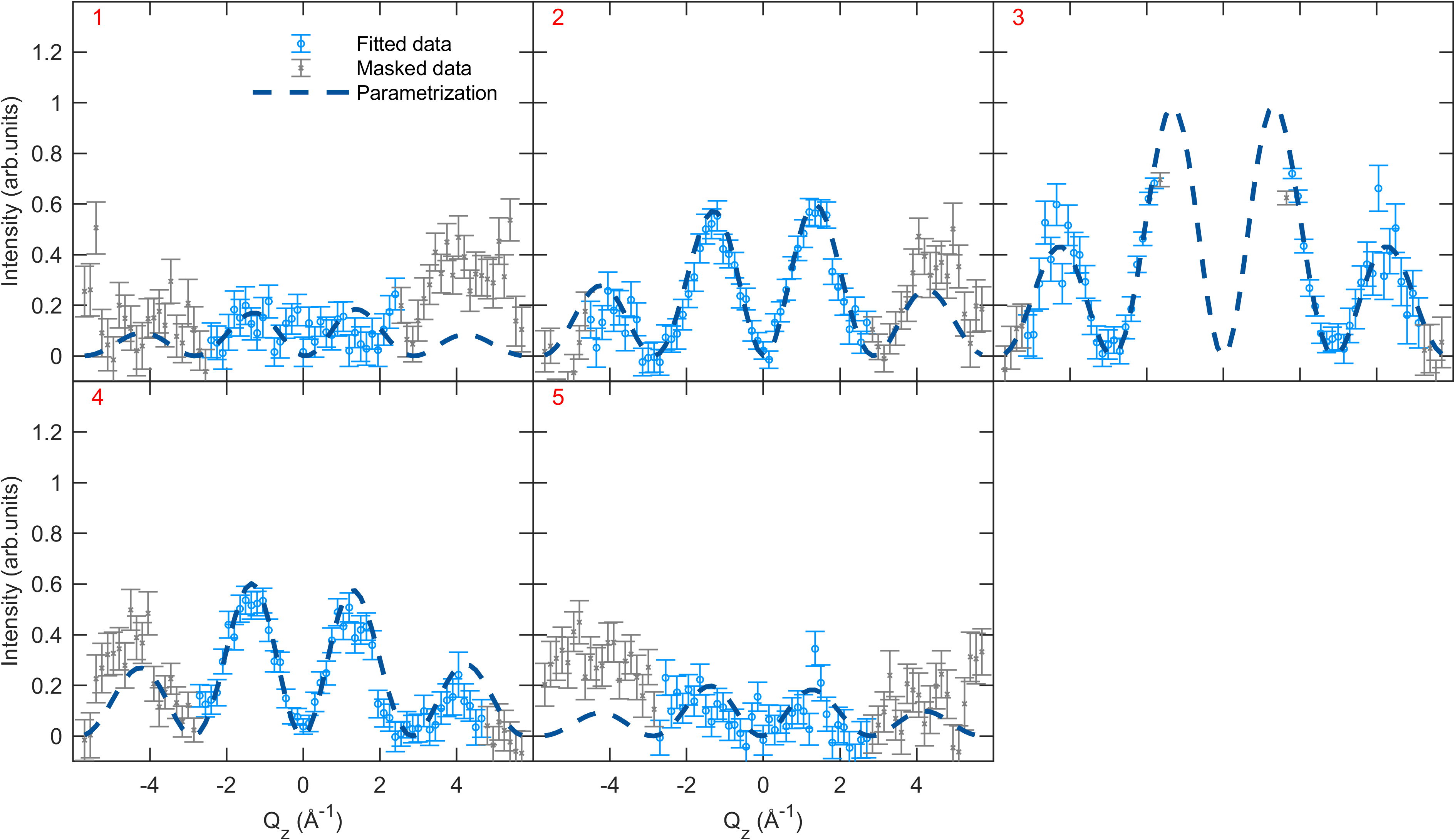}
    \caption{Cuts along the dashed lines in Fig.~\ref{fig:many_cuts_maps}(a). The red number indicates the corresponding cut. Data relevant for the fit are in light blue, whereas those neglected are in black.
}
    \label{fig:many_cuts_oscillation}
\end{figure}

\begin{figure}
    \centering
    \includegraphics[width=1\linewidth]{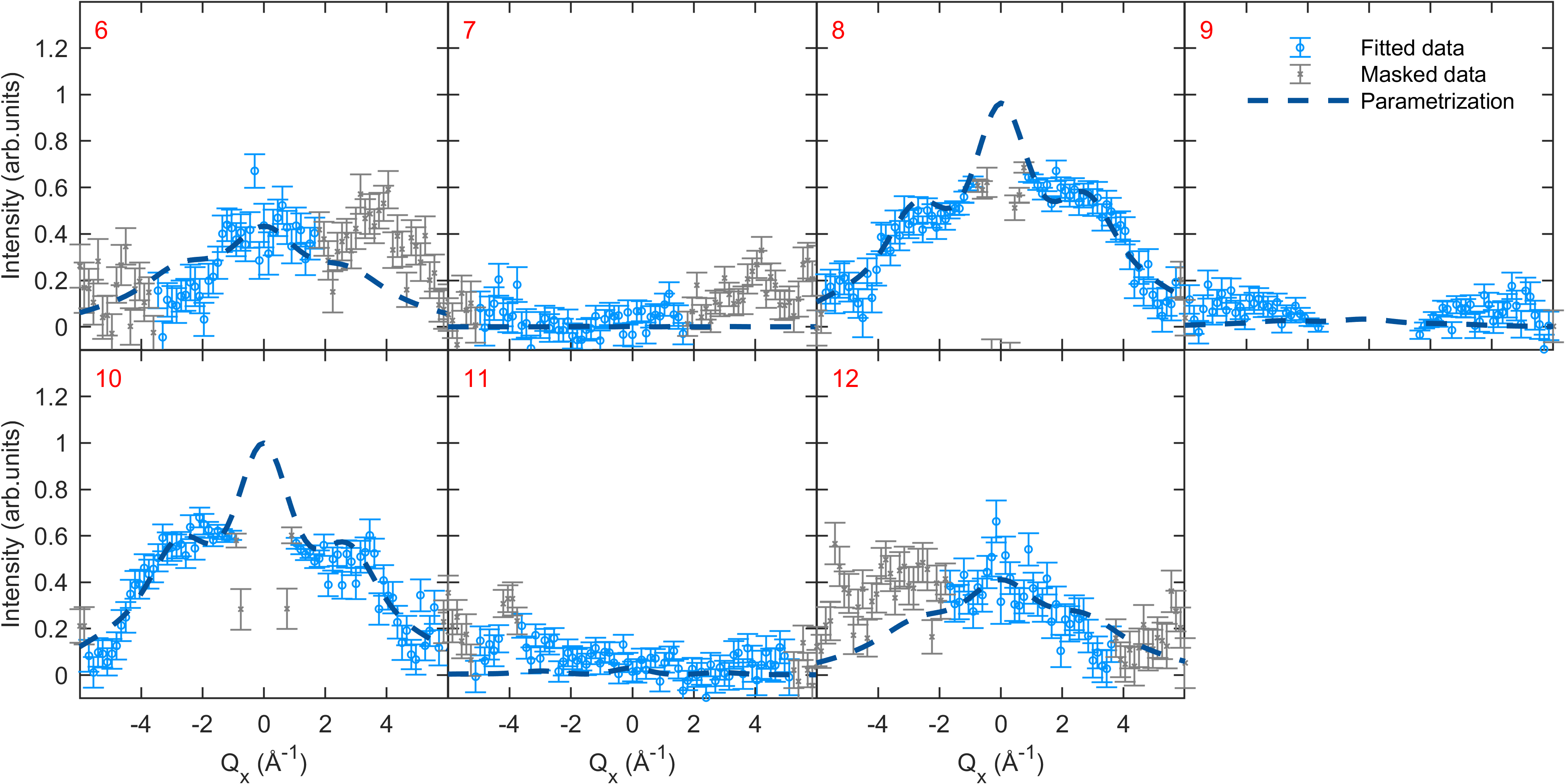}
    \caption{Cuts along the dashed lines in Fig.~\ref{fig:many_cuts_maps}(b). The red number indicates the corresponding cut. Data relevant for the fit are in light blue, whereas those neglected are in black.
}
    \label{fig:many_cuts_ip}
\end{figure}

\begin{figure}
    \centering
    \includegraphics[width=1\linewidth]{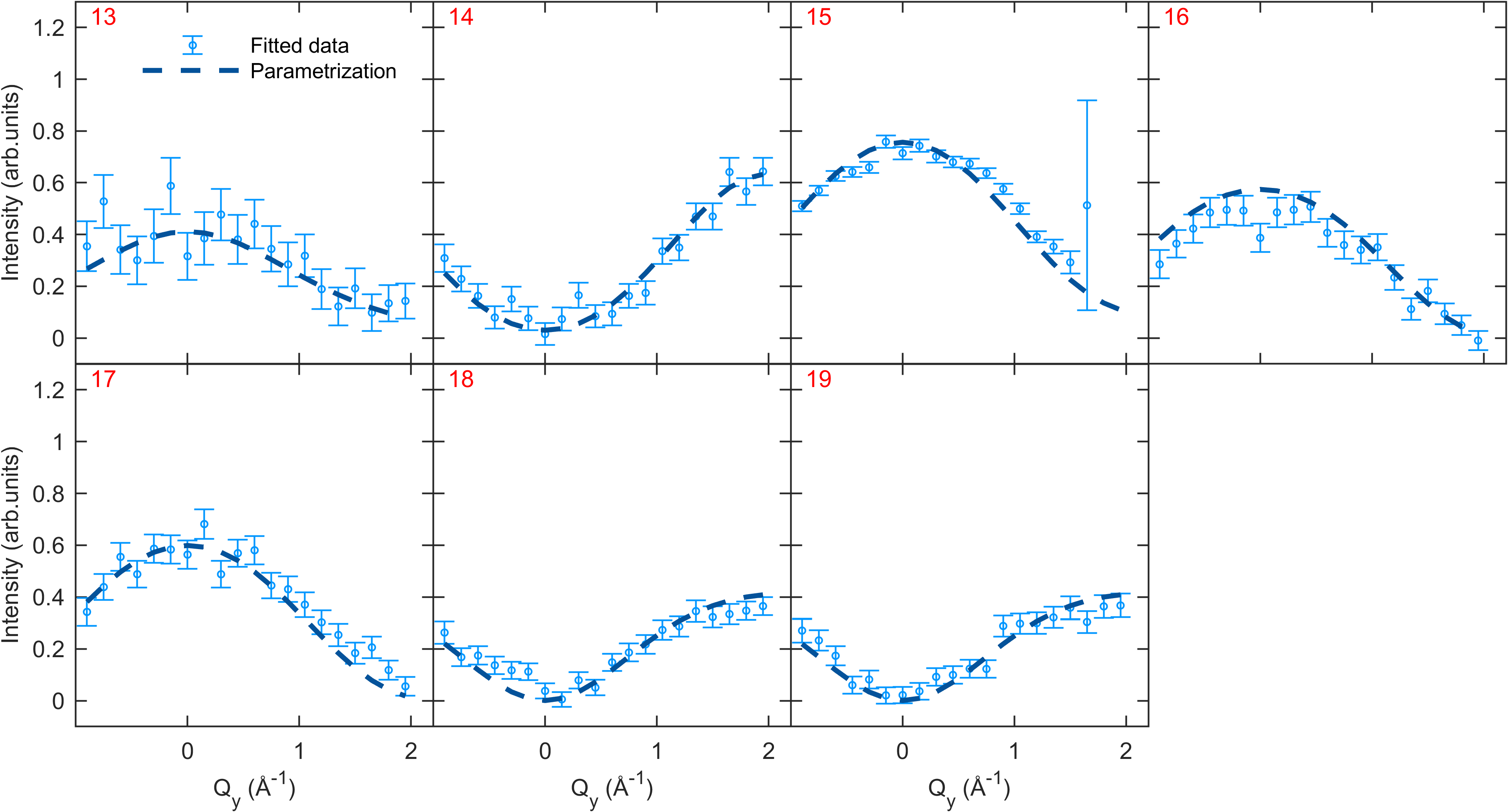}
    \caption{Cuts along the direction perpendicular to the equatorial plane, i.e. $Q_y$, passing through the red diamonds in Fig.~\ref{fig:many_cuts_maps}(c). The red number indicates the corresponding diamond.
}
    \label{fig:many_cuts_oop}
\end{figure}

Figure~\ref{fig:many_cuts_maps} shows the same replica of the magnetic structure factor of the excitation in the equatorial plane. A finite set of line cuts is selected to illustrate the behavior of the analytical parametrization over a broad region of reciprocal space. The dashed lines in Fig.~\ref{fig:many_cuts_maps}(a) probe the direction along the dimer axis in the equatorial plane, and the corresponding line cuts are shown in Fig.~\ref{fig:many_cuts_oscillation}. The dashed lines in Fig.~\ref{fig:many_cuts_maps}(b) probe the direction perpendicular to the cosine modulation, with the corresponding line cuts shown in Fig.~\ref{fig:many_cuts_ip}. Finally, Fig.~\ref{fig:many_cuts_oop} collects a set of line cuts along the $Q_y$ direction, corresponding to the points indicated in Fig.~\ref{fig:many_cuts_maps}(c).

\FloatBarrier

\section{\textit{Ab initio} calculations details}

\subsection{The choice of pseudopotential for spin density calculations}
\begin{figure}
\includegraphics[width=\linewidth]{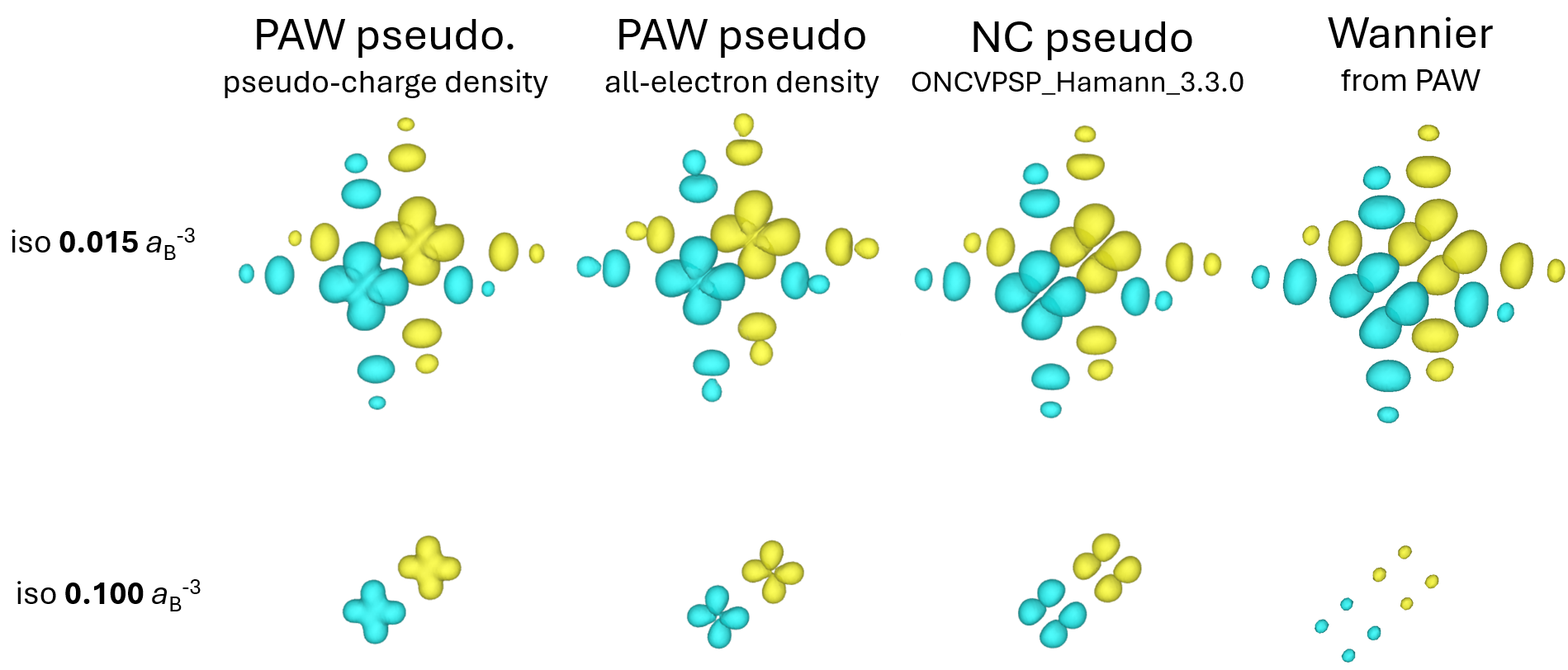}
\caption{Choice of pseudopotential and the resulting spin density displayed for a small (0.015) and large (0.100) value of the electron density isosurface. The PAW pseudopotentials, when used without care, can exhibit a spurious density in the center of the Cu $d$ orbital. Using the all-electron density (\texttt{plot\_num=17} in \texttt{pp.x}) with PAW redeems this issue (see the 0.100 isosurface).}
\label{fig:pseudo_for_spin_density}
\end{figure}

The PAW pseudopotentials, while representing the core orbitals in a more accurate manner, when used by considering only the pseudo-charge density instead of the all-electron density, can result in an incorrect spin density for the central regions; see Fig.~\ref{fig:pseudo_for_spin_density}. The norm-conserving (NC) pseudopotential results are most faithful to the Wannier function density, even though the Wannier functions were obtained for the PAW pseudopotentials.

\subsection{The effective Hubbard \textit{U} from linear response theory}

The Hubbard $U$ for the DFT+$U$ calculations is obtained from linear response theory (LRT)~\cite{cococcioni_linear_2005,timrov_hubbard_2018,timrov_hp_2022} using the \texttt{HP} code and PAW pseudopotentials in a two-step procedure: with a starting point of $U_\mathrm{start}=0.0$~eV we obtain a value of $U_\mathrm{LRT}=8.2$~eV. In a second step, starting from $U_\mathrm{start}=4.5$~eV we obtain a value of  $U_\mathrm{LRT}=7.3$~eV, used throughout the work.

\begin{figure}
\includegraphics[width=0.5\linewidth]{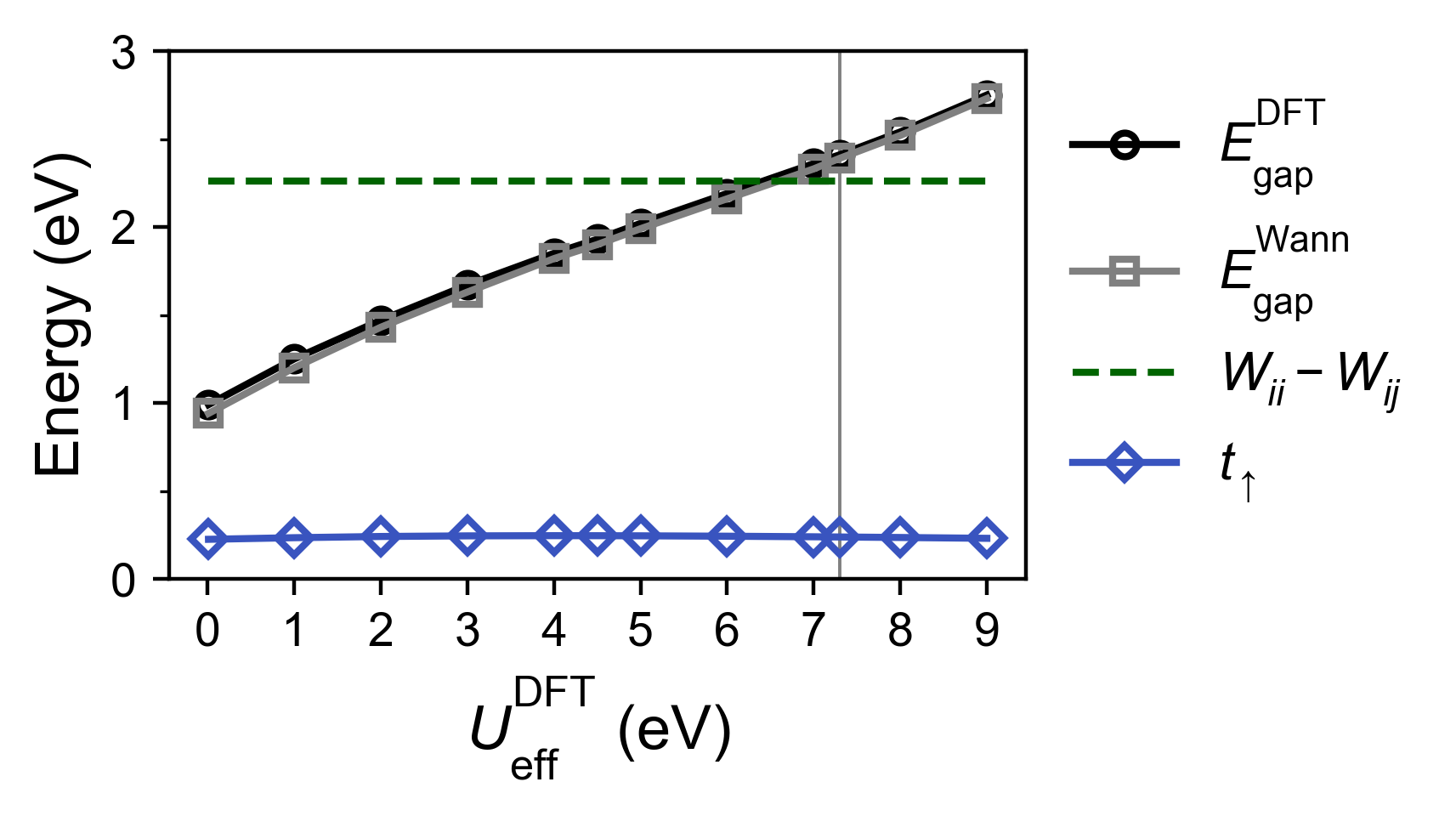}
\caption{The gap between the frontier bands $E_\mathrm{gap}^\mathrm{DFT}$ of the DFT+$U_\mathrm{eff}$ coincides well with the effective gap of the Wannier bands $E_\mathrm{gap}^\mathrm{Wann}$ as well as the gap in the Hubbard dimer model $\approx W_{ii}-W_{ij}$ near the \textit{ab initio}-calculated $U_\mathrm{eff}=7.3$~eV. The hopping integral $t\approx 0.24$~eV is almost constant with $U_\mathrm{eff}$.}
\label{fig:Hubbard_gap_vs_Hubbard_U}
\end{figure}

In Figure~\ref{fig:Hubbard_gap_vs_Hubbard_U} we show the evolution of the effective gap between the frontier Hubbard bands from the DFT, Wannier bands and the Hubbard dimer model values, which all coincide near the \textit{ab initio}-calculated $U_\mathrm{eff} = 7.3$~eV, confirming the validity of this choice.

\subsection{cRPA calculations}
\begin{figure}
\includegraphics[width=0.75\linewidth]{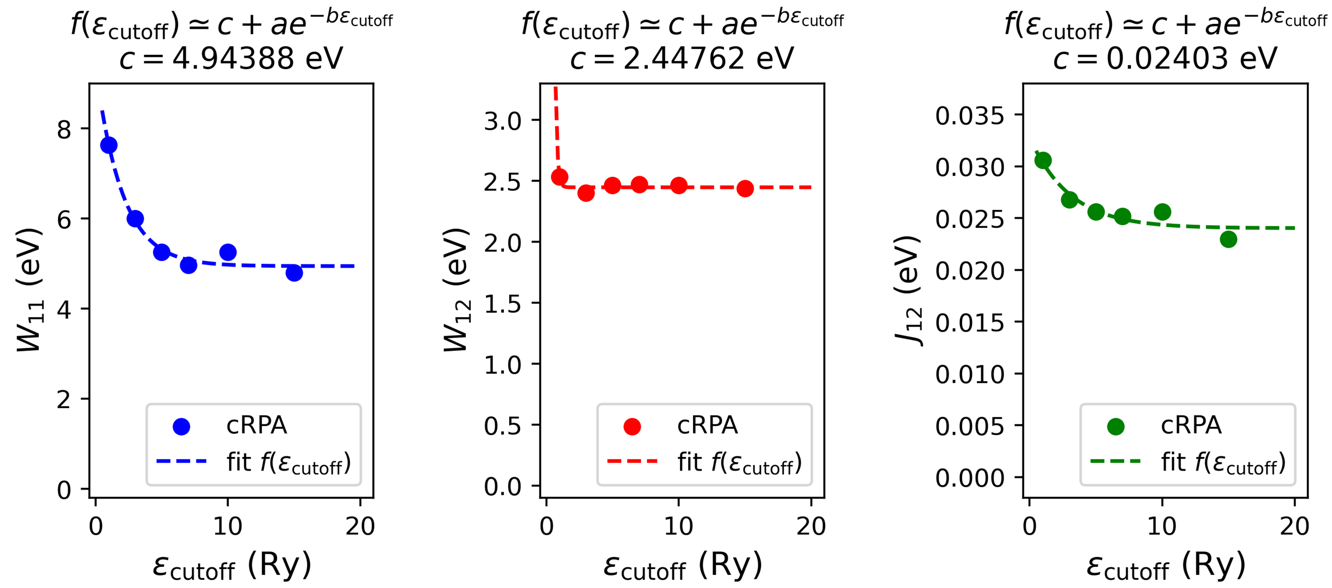}
\caption{Convergence of the 4-body integrals with respect to the cutoff energy for the dielectric matrix $\varepsilon_\mathrm{cutoff}$.}
\label{fig:cRPA_convergence}
\end{figure}

We converge the screened 4-point integrals computed by \texttt{RESPACK}~\cite{nakamura_respack_2021} within cRPA with respect to the cutoff energy for the dielectric matrix $\varepsilon_\mathrm{cutoff}$  by fitting an exponential $W_{ij}(\varepsilon_\mathrm{cutoff}) = c + a \cdot e^{-b \varepsilon_\mathrm{cutoff}}$ and identifying the converged $W_{ij}$ with the fitted $c$ parameter, as shown in Fig.~\ref{fig:cRPA_convergence}.

\end{document}